\documentclass[12pt,a4paper]{article}

\usepackage{setspaceff}
\def\bodystretch{1.20}
\setfootnotestretch{\bodystretch}
\setfloatstretch{\bodystretch}

\usepackage{curves}
\usepackage{ifthen}
\newboolean{hi_s}   
\newboolean{lo_s}

\usepackage{epsfig}

\usepackage{amssymb}
\usepackage{amsmath}

\usepackage{color}

\date{
\pp
\today
\vspace{-4ex}
}
\title{
\sc
Undiscounted Bandit Games\thanks{
We thank three anonymous referees and an advisory editor for their comments.
Our thanks for helpful discussions and suggestions are owed to
Chris Harris,
Thomas Kesselheim,
Alois Kneip,
Albert N.\ Shiryaev,
Philipp Strack,
Bruno Strulovici,
Bernhard von Stengel,
and
seminar participants in
Austin,
Budapest,  
Cambridge,
Florence,
Keele,
Oxford,
Paris,
Shanghai,
Southampton,
and
UCL/Birkbeck.
We thank
the Center for Economic Studies at the University of Munich
and
the Studienzentrum Gerzensee
for their hospitality.
Financial support from the Deutsche Forschungsgemeinschaft through GRK 801, SFB TR 15 (Project A08) and SFB TR 224 (Project B04) is gratefully acknowledged.
}}
\author{
Godfrey Keller\thanks{
Department of Economics, University of Oxford,
Manor Road Building, Oxford OX1 3UQ, UK.
}
\and
Sven Rady\thanks{
Department of Economics and Hausdorff Center for Mathematics, University of Bonn,
Adenauerallee 24-42, D-53113 Bonn, Germany.
}
}

\let\bodyenvironmentsize=\normalsize
\let\bodyproofsize=\normalsize
\let\appenvironmentsize=\small
\let\appproofsize=\small

\newcommand\AppendixIn{     
    \newpage
    \section*{Appendix}

    \parskip0.0ex
    \let\environmentsize=\appenvironmentsize
    \let\proofsize=\appproofsize
    \environmentsize

    \renewcommand{\thesection}{\Alph{section}}      
    \renewcommand{\theequation}{\thesection.\arabic{equation}}  

    \setcounter{section}{0}
}
\newcommand\AppendixOut{    
    \renewcommand{\thesection}{\arabic{section}}    
    \renewcommand{\theequation}{\arabic{equation}}   

    \parskip1.5ex
    \let\environmentsize=\bodyenvironmentsize
    \let\proofsize=\bodyproofsize
    \environmentsize
}

\newtheorem{lemma}{Lemma}[section]
\newtheorem{cor}  {Corollary}[section]

\newcommand{\proof}      [1]{\proofsize
        {\sc Proof:}\ #1 \ \QED
        \par \environmentsize \vskip8truept}

\def\QED{\hfill \rule{3mm}{3mm}}

\def\p   {\vskip4truept\noindent}
\def\pp  {\vskip8truept\noindent}

\newcommand{\astrut}[1]{\rule{0em}{#1ex}}

\newcommand{\eref}[1]{{\rm (\ref{#1})}} 
\newcommand{\divn}[2]{\mbox{$\frac{#1}{#2}$}}
\def\half{\divn{1}{2}}

\def\Pr  {\mathbb{P}} 
\def\Exp {\mathbb{E}} 
\def\Var {\mbox{Var}}
\def\Cov {\mbox{Cov}}

\def\R   {\mathbb{R}} 
\def\Rnotz{\R\!\setminus\!\{0\}}
\def\S   {\mathbb{S}}

\begin{document}
\maketitle
\setcounter{page}{0}
\thispagestyle{empty}

\begin{abstract}

We analyze undiscounted continuous-time games of strategic experimentation with two-armed bandits.
The risky arm generates payoffs according to a L\'{e}vy process with an unknown average payoff per unit of time which nature draws from an arbitrary finite set.
Observing all actions and realized payoffs, plus a free background signal, players use Markov strategies with the common posterior belief about the unknown parameter as the state variable.
We show that the unique symmetric Markov perfect equilibrium can be computed in a simple closed form involving only the payoff of the safe arm, the expected current payoff of the risky arm, and the expected full-information payoff, given the current belief.
In particular, the equilibrium does not depend on the precise specification of the payoff-generating processes.
\pp
\p
{\em JEL} {\sc Classification Numbers:}
C73, 
D83. 

\end{abstract}

\newpage

\AppendixOut                


\section{Introduction}

We analyze a class of continuous-time two-armed bandit models in which a number of symmetric players act non-cooperatively, trying to learn an unknown state of the world that governs the risky arm's expected payoff per unit of time.
Actual payoffs are given by L\'{e}vy processes, that is, processes with independent and stationary increments.
In addition, players receive free background information in the form of a process of the same type as the payoff processes.
Rather than discounting future payoffs, players evaluate their payoff streams according to the strong long-run average criterion.\footnote{
First used by Ramsey (1928) in growth theory, this criterion is the limit of the standard discounted performance criterion as the discount rate goes to zero, both in terms of value functions and optimal strategies.
See Dutta (1991) for the connection between performance criteria with and without discounting in discrete time, and
Bolton and Harris (2000) for a detailed treatment of the strong long-run average criterion in a continuous-time Bayesian-learning setting such as ours.}
Assuming that all actions and payoffs are public information, we restrict players to Markov strategies with the common posterior belief about the unknown parameter as the natural state variable, and we characterize the unique symmetric Markov perfect equilibrium.

This setting allows us to handle a much larger class of priors and payoff-generating processes than the existing economics literature on bandit-based multi-agent learning in continuous time.
First, the unknown state of the world can be drawn from an arbitrary finite set, whereas the literature assumes a binary state.
Second, the payoff processes can combine continuous with discrete increments, whereas the literature assumes either Brownian or Poisson payoffs.
Third, lump-sum payoffs can be good or bad news, whereas the literature assumes that news is of one type only.

Despite this broadening of the class of payoff-generating processes,
and the generalization from Bernoulli to arbitrary discrete priors in particular,
the equilibrium strategy has a simple explicit form.
In fact, at any point in time, a player's best response depends only on the intensity of experimentation performed by the other players, the payoff of the safe arm, the expected current payoff of the risky arm, and the expected full-information payoff
-- it does \emph{not} depend on the precise specification of the payoff-generating process.
This feature carries over to the symmetric Markov perfect equilibrium, where one and the same functional form applies across all specifications that we consider.
The common equilibrium action is a piecewise linear function of the ratio of two differences: that between the risky arm's expected full-information payoff and the safe payoff, and that between the safe payoff and the risky arm's expected current payoff.

We further show that this result extends to two specifications of priors and payoff-generating processes in which the unknown state of the world is drawn from a continuous distribution of unbounded support:
Brownian payoffs with normal priors, and Poisson payoffs with gamma priors.
In either specification, the players' information is captured by a two-dimensional sufficient statistic, which can serve as the state variable for Markov strategies.

Our characterization of the unique symmetric Markov perfect equilibrium hinges on four features of the settings that we study:
(i) players receive free background information;
(ii) they use the strong long-run average criterion;
(iii) the experimentation game is played in continuous time;
and (iv) the players' risky payoff processes and the background information are all of the same (unknown) type, hence perfect substitutes with respect to learning.

In fact, the background information ensures that players learn the true state exponentially fast, no matter what strategy profile they use.\footnote{
The assumption of free background information is quite natural in a number of situations in which players face trade-offs and incentives akin to those captured in the experimentation game at hand.
Examples are experience goods (e.g.\ the choice between a familiar restaurant of known quality and a new restaurant of unknown quality) and innovation in research teams.
Consumers have access to guides or online recommender systems, research teams to published patents or journal articles.}
This makes it possible to evaluate players' random payoff streams according to the strong long-run average criterion, that is, by computing the expected accumulated shortfall of realized payoffs relative to the expected full-information payoff.
Under this criterion, the problem of finding a best response to the opponents' Markovian strategy profile has a recursive structure amenable to dynamic-programming techniques.
In continuous time, this leads to a Hamilton-Jacobi-Bellman (HJB) equation in which the value function (i.e.\ the payoff function induced by a best response) enters only through the expected rate of change of continuation payoffs.
When the players' risky payoff processes and the background information are all of the same type, moreover, the expected rate of change of continuation values is linear in the total intensity of experimentation.
This makes it possible to eliminate a player's value function completely from the maximization problem in the HJB equation, so optimal actions can be determined belief by belief without reference to the value function and the payoff-generating processes.

While the computation of these candidate best responses does not involve the specifics of the payoff-generating processes, the evolution of the players' posterior beliefs obviously does depend on them, and it is essential to ensure that the strategies in question induce a well-defined law of motion for those beliefs.
Standard results on the existence and uniqueness of solutions to stochastic differential equations require Lipschitz continuity of coefficients.
We restrict players to strategies that are Lipschitz continuous in the posterior belief, therefore, and we show that the candidate symmetric equilibrium strategy obtained from the HJB equation falls in this class.

To verify that all players using this strategy constitutes an equilibrium, we exploit the fact that a player's value function is the unique viscosity solution of the HJB equation subject to the relevant boundary conditions.\footnote{
For an introduction to viscosity solutions in continuous-time stochastic control with applications in economics and finance, see the textbooks by {\O}ksendal and Sulem (2007, Chapter 9) and Pham (2009, Chapter 4).}
By showing that the payoff function for the suggested strategy profile also solves this boundary-value problem in the viscosity sense, we establish that the two functions agree, so the player indeed plays a best response.\footnote{
For a recent application of the viscosity-solutions approach to the verification of optimality in a single-agent learning context with Brownian signals, see Ke and Villas-Boas (2019).}

This paper belongs to a large and still growing economics literature, surveyed by Bergemann and V\"{a}lim\"{a}ki (2008) and H\"{o}rner and Skrzypacz (2016), that analyzes Bayesian bandit problems in contexts such as search, learning, pricing, matching, contracting and information design.
More specifically, we contribute to the literature on strategic experimentation with bandits, sharing its most basic setup -- symmetric players solving identical bandit problems with a safe and a risky arm, and exerting a purely informational externality on each other.
This literature was initiated by Bolton and Harris (1999) who characterize the unique symmetric Markov perfect equilibrium under discounting when risky payoffs are generated by Brownian motions with an unknown drift that can be either high or low.
The equilibrium features free-riding on other players' experimentation efforts, but also an encouragement effect whereby future experimentation by others increases a player's current effort.

Studying the same setting under the strong long-run average criterion, and adding background information, Bolton and Harris (2000) are able to characterize the entire set of undiscounted Markov equilibria -- symmetric and asymmetric.
This is possible because posterior beliefs in the binary Brownian model evolve as a one-dimensional diffusion, which allows for a space of admissible Markov strategies large enough to accommodate the discontinuities of actions with respect to beliefs which are an immutable feature of asymmetric equilibria.

Keller, Rady and Cripps (2005) and Keller and Rady (2010, 2015) analyze symmetric and asymmetric Markov perfect equilibria under discounting (and without background information) when the payoffs are generated by Poisson processes with an unknown intensity that can be either high or low.
Here, it is the piecewise deterministic evolution of beliefs on the unit interval that permits discontinuous strategies:
the belief moves in one and the same direction as long as no lump-sum payoff arrives, so one-sided continuity of strategies suffices to induce well-defined belief dynamics.

Our restriction to Lipschitz continuous strategies, mandated by the more general payoff-generating processes and priors that we consider, rules out asymmetric equilibria but, as our main result shows, still permits the computation of a unique symmetric equilibrium along the same lines, and of the same functional form, as in Bolton and Harris (2000).
The strategic forces at work are also the same:
free-riding on the information produced by other players and the background signal, but no encouragement effect because players do not discount.
We bring two entirely new elements to this analysis, though: a proof of exponentially fast convergence of beliefs which in turn implies boundedness of the strong long-run average criterion, and the use of the viscosity-solutions approach in the verification of the best-response property.

Relative to the above body of work, our main contribution is to show that the use of the strong long-run average criterion and the introduction of background information permit the computation of a unique symmetric Markov perfect equilibrium in a simple explicit form for a much broader class of payoff-generating processes and priors.
In particular, this allows the analysis of situations in which players learn both from a Brownian payoff component that might capture the steady flow of information under `business as usual' and from a jump component that might capture the sudden bursts of information -- good and bad -- arriving in exceptional times or `crises'.\footnote{
Cohen and Solan (2013) analyze a single-agent version of the bandit problem with L\'{e}vy payoffs in which the state is binary and jumps are always good news; assuming that players discount the future, they can dispense with background information.}
In our framework, nature could draw the characteristics of each component from separate (finite) sets, moreover, with full flexibility as to their joint distribution.

A second contribution is to delineate precisely which aspects of the analysis in Bolton and Harris (2000) and the present paper generalize to other settings and which do not.
In fact, each of the four features mentioned above is crucial.
First, without background information, the strong long-run average criterion would have no power because the expected accumulated shortfall of received payoffs relative to the expected full-information payoff would grow infinitely large in general.
Second, with discounting, the HJB equation would necessarily contain a term `discount rate times current value' that is not multiplied by the total intensity of experimentation, so best responses would depend on current values.
As pointed out in Dutta (1991), moreover, alternative undiscounted performance criteria would not permit a recursive representation.
Third, if the model were set in discrete time, the expected rate of change of continuation payoffs would not be linear in the total intensity of experimentation.
In a discrete-time version of the game in Keller, Rady and Cripps (2005), for example, the probability of a success in any given round is clearly non-linear in the number of players pulling the risky arm.
Fourth, linearity would also fail if the type of the risky arm were independent or imperfectly correlated across players,
if the law of the payoff process differed across players, or if each player had access to more than one risky arm.\footnote{
Linearity would also fail in a restless bandit model in which the state of the world changed exogenously over time.
This would be the case, for example, if payoffs were generated by a Brownian motion with an unknown drift subject to Markovian state-switching between a high and a low level as in Keller and Rady (1999, 2003).}

Besides Bolton and Harris (2000), the undiscounted limit of a continuous-time stochastic game with a one-dimensional diffusion state has also been studied in Harris (1988, 1993) and Bergemann and V\"{a}lim\"{a}ki (1997, 2002).
More recent applications of this methodology to single-agent experimentation problems can be found in Bonatti (2011) and Peitz, Rady and Trepper (2017).
The above considerations may be useful for applications of the strong long-run average criterion in continuous-time stochastic optimization problems and games with richer state spaces and dynamics.

Through its use of the strong long-run average criterion, which permits a minimization-of-regret interpretation, the paper is loosely related to the vast literature on no-regret learning in non-Bayesian bandit problems or games; see Blum and Mansour (2007) or Bubeck and Cesa-Bianchi (2012), for example.
While the emphasis in this literature is on upper bounds for cumulative regret and on the speed of convergence to an optimum or Nash equilibrium, the strategic-experimentation literature approaches the entire learning process from a dynamic equilibrium perspective -- that is, it analyzes learning \emph{in}, not \emph{of}, an equilibrium.

The rest of the paper is organized as follows.
Section \ref{sec: model} sets up the game and states our assumptions on priors, payoff-generating processes and strategy spaces.
Section \ref{sec: gen} presents the infinitesimal generator of the process of posterior beliefs.
Section \ref{sec: equil} constructs the unique symmetric Markov perfect equilibrium and discusses its properties.
Section \ref{sec: examples} presents extensions of our analysis to two settings with a continuously distributed state of the world.
Section \ref{sec: concl} offers some concluding remarks.


\section{The Experimentation Game
\label{sec: model}}

Time $t \in [0,\infty[$ is continuous.
There are $N \geq 1$ players, each of them endowed with one unit of a perfectly divisible resource per unit of time.
Each player faces a two-armed bandit problem where she continually has to decide what fraction of the available resource to allocate to each arm.
One arm is safe, the other risky.

The safe arm generates a known constant payoff $s > 0$ per unit of time.
The evolution of the payoffs generated by the risky arm depends on a state of the world, $\ell$, which nature draws from the set $\{0,1,\ldots,L\}$ with $L \geq 1$ according to the positive probabilities $\pi_0,\ldots,\pi_L$.
Players do not observe the state, but know its distribution.
They also know that the payoff process associated with player $n$'s risky arm is of the form
$$
X^n_t = \rho\,t+\sigma Z^n_t + Y^n_t,
$$
where
$Z^n$ is a standard Wiener process
and
$Y^n$ is a compound Poisson process whose L\'{e}vy measure $\nu$ is finite and has a finite second moment $\int h^2\,\nu(dh)$.\footnote{
Here, $\nu(B) < \infty$ is the expected number of jumps per unit of time whose size is in the Borel set $B \subseteq \Rnotz$.
The finite second moment ensures that the processes $X^n$ have finite mean and finite quadratic variation.
A reminder about L\'{e}vy processes can be found in Cohen and Solan (2013, Section 2.1).}
The drift rate $\rho$, the diffusion coefficient $\sigma >0$ and the L\'{e}vy measure $\nu$ are the same for all players.
While $\sigma$ is the same in all states of the world, moreover,
$\rho$ and $\nu$ vary with the state.\footnote{
Our assumptions on the diffusion coefficient and the L\'{e}vy measures ensure that the players cannot infer the true state instantaneously from the continuous and jump part of risky payoffs, respectively.
Finiteness of L\'{e}vy measures simplifies the exposition but can be dropped: it suffices that these measures have a finite second moment and satisfy assumptions A3 and A4 of Cohen and Solan (2013) for each pair of states.
Note also that the framework is flexible enough to accommodate drift rates and L\'{e}vy measures that are drawn separately from some finite set each.
The states $0,1,\ldots,L$ then correspond to the possible pairs $(\rho,\nu)$, and the probabilities $\pi_0,\ldots,\pi_L$ describe their joint distribution.}
Conditionally on $\ell$, the processes $Z^1,\ldots,Z^N,Y^1,\ldots,Y^N$ are independent.

We write $\rho_\ell$ and $\nu_\ell$ for the drift rate and L\'{e}vy measure in state $\ell$,
$\lambda_\ell = \nu_\ell(\Rnotz)$ for the expected number of jumps per unit of time,
and $h_\ell = \int_{\Rnotz} h\,\nu_\ell(dh)\,/\,\lambda_\ell$ for the expected jump size.

The state-contingent expected risky payoff per unit of time is $\mu_\ell = \rho_\ell + \lambda_\ell \, h_\ell$.
We assume that $\mu_0 < \mu_1 < \ldots < \mu_{L-1} < \mu_L$ with $\mu_0 <  s < \mu_L$, so that neither arm dominates the other in terms of expected payoffs.
Writing $\pi$ for the vector of probabilities $(\pi_1,\ldots,\pi_L)$, we let $m(\pi)$ denote the expected per-period payoff from the risky arm,
and $f(\pi)$ a player's expected per-period payoff under full information:\footnote{
Given our convention to summarize the distribution of the unknown state by the probabilities of the realizations $1,\ldots,L$, the symbol $\pi_0$ should be viewed as shorthand for $1-\sum_{\ell=1}^L \pi_\ell$ from now on.}
$$
  m(\pi) = \sum_{0}^{L} \pi_\ell \mu_\ell,
\qquad
  f(\pi) = \sum_{0}^{L} \pi_\ell \max\{s, \mu_\ell\}.
$$

Let $k_{n,t} \in [0,1]$ be the fraction of the available resource that player $n$ allocates to the risky arm at time $t$; this fraction is required to be measurable with respect to the information that the player possesses at time $t$.
The player's cumulative payoff up to time $T$ is then given by the time-changed process
$\left[ T-\tau^n(T) \right] s + X^n_{\tau^n(T)}$
where
$\tau^n(T)=\int_0^T k_{n,t} \, dt$
measures the \emph{operational time} that the risky arm has been used.
As $X^n_t - \mu t$ is a martingale, the player's expected payoff up to $T$ 
is
$$\Exp\left[\int_0^T \{(1-k_{n,t})s + k_{n,t} \mu\} \,dt\right];$$
here, the expectation is both about the process of allocations $k_{n,t}$ and the unknown expected per-period payoff $\mu$.
With $s$ lying in the interior of the range of possible realizations of $\mu$,
each player has an incentive to learn the quality of the risky arm.

Players do not discount future payoffs;
as in Bolton and Harris (2000), they are instead assumed to use the strong long-run average criterion.
This means that player $n$ chooses allocations $k_{n,t}$ so as to maximize
$$
  \Exp \left[ \int_0^\infty
         \left\{\astrut{2} (1-k_{n,t})s + k_{n,t} m(\pi) - f(\pi) \right\}\,dt
         \right].
$$
Here, the integrand is the difference between what a player expects to receive at a given point in time and what she would expect to receive were she to be fully informed.\footnote{
The integral is the negative of the limit as $T \to \infty$ of
$
T f(\pi) - \Exp \left[ \int_0^T
         \left\{\astrut{2} (1-k_{n,t})s + k_{n,t} m(\pi) \right\}\,dt
         \right]\!,
$
the continuous-time equivalent of the expected regret (or `Bayes risk') considered in Lai (1987), for example.
In the present context, the strong long-run average criterion is equivalent to minimization of cumulative Bayesian regret, therefore.}
Note that this objective function depends on others' actions only through their impact on the player's own choices.
In fact, we will soon impose restrictions under which others' actions matter only through their effect on a player's beliefs.

The players start with a common prior belief about the unknown state $\ell$, given by the probabilities with which nature draws this state.
Thereafter, all observe each other's actions and outcomes as well as a common background signal,
so they hold common posterior beliefs throughout time.
The background signal is generated by the time-changed process $X^0_{\tau^0(t)}$ where $X^0$ is an independent process of the same law as each player's payoff process from the risky arm, and $\tau^0(t) = k_0 t$ with $k_0 > 0$ exogenously given and arbitrarily small.
This signal ensures that the players eventually learn the value of $\mu$ even if they all play safe all the time.

Let $\pi_t$ denote the vector of common posterior probabilities that the players assign to states $1, \ldots, L$ given their observations up to time $t$.
With respect to the information filtration generated by these observations, the process of beliefs $\pi_t$ is a Markov process (in fact, a jump diffusion) and a martingale.
The linearity of the functions $m$ and $f$ now implies that $\Exp[m(\pi_t)] = m(\pi)$ and $\Exp[f(\pi_t)] = f(\pi)$ for all $t > 0$,
so we can rewrite the above objective function as
$$
  \Exp \left[ \int_0^\infty
         \left\{\astrut{2} (1-k_{n,t})s + k_{n,t} m(\pi_t) - f(\pi_t) \right\}\,dt
       \right],
$$
highlighting the potential for the posterior belief to serve as a state variable.

From now on, we restrict players to strategies that are Markovian with respect to this variable, so that the action $k_{n,t}$ chosen at time $t$ is a deterministic function of $\pi_t$ only.\footnote{
In the presence of discrete payoff increments, one actually must take the left limit $\pi_{t-}$ as the state variable because the action chosen at time $t$ cannot depend on a lump-sum payoff that arrives at $t$.
We write $\pi_t$ with the understanding that the left limit is meant whenever this distinction is relevant.}
More precisely, we take the players' common strategy space ${\cal K}$ to be the set
of all Lipschitz continuous functions from the $L$-dimensional simplex
$$\Delta_L = \left\{ \pi \in \R^L_+\!: \sum_{\ell =1}^L \pi_\ell \leq 1 \right\}$$
to $[0,1]$.
By standard existence and uniqueness results for solutions of stochastic differential equations, any strategy profile $(\kappa_1,\ldots,\kappa_N) \in {\cal K}^N$ gives rise to a well-defined process of posterior beliefs,\footnote{
For $L=1$ and no discontinuous payoff component, i.e.\ in the setting analyzed in Bolton and Harris (2000), the presence of background information allows one to invoke a result of Engelbert and Schmidt (1984) whereby any profile of Borel measurable Markov strategies implies a unique solution for the belief dynamics; see also Section 5.5 of Karatzas and Shreve (1988).
For $L=1$, no Brownian payoff component, and lump-sum payoffs that are always good news (meaning that $\nu_0(B) \leq \nu_1(B)$ for all Borel sets $B \subseteq \Rnotz$), one can proceed as in Keller, Rady and Cripps (2005) and Keller and Rady (2010) and take $\cal K$ to be the set of functions which are left-continuous and piecewise Lipschitz continuous; as beliefs drift down deterministically in between lump-sums, these properties allow one to construct belief dynamics in a pathwise fashion.
Neither approach generalizes to higher dimensions.}
and hence to well-defined payoffs
$$
  u_n(\pi|\kappa_1,\ldots,\kappa_N)
  =  \Exp \left[ \left. \int_0^\infty
         \left\{\astrut{2} [1-\kappa_n(\pi_t)] s + \kappa_n(\pi_t) m(\pi_t) - f(\pi_t) \right\}\,dt
         \ \right| \pi_0 = \pi \right] \in [-\infty,0].
$$

A player's payoff will indeed be $-\infty$ for certain Markov strategies.
If the player always uses the safe arm, for example, and the true state $\ell$ is such that $\mu_\ell > s$,
then by almost sure convergence of posterior beliefs to the truth, the above integrand will converge to $s - \mu_\ell < 0$ as $t$ grows large,
implying a diverging integral in that state.
Since this occurs with positive prior probability, the expected payoff is $-\infty$, therefore.

The following considerations lead to a class of strategies with finite expected payoffs.
Let $\Delta^1_L$ be the set of beliefs $\pi \in \Delta_L$ such that $\pi_\ell = 0$ whenever $\mu_\ell < s$,
and $\Delta^0_L$ the set of beliefs $\pi \in \Delta_L$ such that $\pi_\ell = 0$ whenever $\mu_\ell > s$.
These sets are the faces of the simplex on which there is a trivially optimal action: the risky arm on $\Delta^1_L$, and the safe arm on $\Delta^0_L$.
We call a strategy $\kappa_n \in \cal K$ \emph{reasonable} if $\Delta^1_L$ is contained in the interior of $\kappa_n^{-1}(1)$ and $\Delta^0_L$ is contained in the interior of $\kappa_n^{-1}(0)$.
Given such a strategy, each face of the simplex on which there is a trivially optimal action has a neighbourhood in $\Delta_L$ on which the strategy selects that action and thus maximizes the expected per-period payoff, so that $[1-\kappa_n(\pi)] s + \kappa_n(\pi) m(\pi) = \max\{s,m(\pi)\}$.
On $\Delta^1_L$ and $\Delta^0_L$, moreover, $\max\{s,m(\pi)\} = f(\pi)$, hence
$[1-\kappa_n(\pi)] s + \kappa_n(\pi) m(\pi) - f(\pi) = 0$;
this holds in particular at each vertex of the simplex.
Establishing that, in the presence of background information, posterior beliefs converge exponentially fast to the truth, we show in the appendix that the expected payoff from a reasonable strategy is always finite and, in fact, bounded on the simplex, irrespective of the opponents' profile of Markov strategies.\footnote{
In the minimization-of-regret interpretation of the payoff criterion, this means that all reasonable strategies have zero time-averaged Bayesian regret:
$
f(\pi) - \frac{1}{T }\Exp \left[ \int_0^T
         \left\{\astrut{2} (1-k_{n,t})s + k_{n,t} m(\pi) \right\}\,dt
         \right] \to 0
$
as $T \to \infty$.
No-regret learning in this specific sense is thus strictly less restrictive than the strong long-run average criterion which, as we shall see, selects a unique optimal strategy given opponents' play.}

Strategy $\kappa_n \in {\cal K}$ is a \emph{best response} against $\kappa_{\neg n}=(\kappa_1,\ldots,\kappa_{n-1},\kappa_{n+1},\ldots,\kappa_N) \in {\cal K}^{N-1}$
if $u_n(\pi|\kappa_n,\kappa_{\neg n}) \geq u_n(\pi|\tilde{\kappa}_n,\kappa_{\neg n})$ for all $\pi \in \Delta ^L$ and all $\tilde{\kappa}_n \in {\cal K}$.
A \emph{Markov perfect equilibrium (MPE)} is a profile of strategies $(\kappa_1,\ldots,\kappa_N) \in {\cal K}^N$ that are mutually best responses.
Such an equilibrium is \emph{symmetric} if $\kappa_1 = \kappa_2 = \ldots = \kappa_N$.
Obviously, each player must obtain a finite payoff in any MPE.


\section{The Infinitesimal Generator
\label{sec: gen}}

The evolution of posterior beliefs is driven by up to $N+1$ distinct sources of information: the observations on up to $N$ risky arms plus the background signal.
Suppose that only player 1 uses the risky arm, and at full intensity.
In other words, consider the time-invariant action profile for which $k_1 = 1$ whereas $k_n = 0$ for all $n > 1$.
Write ${\cal G}$ for the infinitesimal generator of the corresponding belief process -- as the payoff-generating process is the same on every player's risky arm, the identity of the player in question does indeed not matter here.

If we now change player 1's time-invariant intensity to $k_1 < 1$ while keeping all other intensities at zero, the resulting deceleration of the process of observations implies the scaled-down generator $k_1 \cal G$ for the posterior belief; see Dynkin (1965, Theorem 10.12), for example.
The same applies to the background signal, of course, if it alone is observed, with associated generator $k_0 \cal G$.

As the processes $X^0$ and $X^1$ are independent conditionally on the realized state, Trotter (1959, Theorem 1)
implies that the infinitesimal generator of posterior beliefs is $(k_0 + k_1) \cal G$ when both the background signal and player 1's payoffs are observed.
By the same token, successively adding the other players with time-invariant allocations $k_2,\ldots,k_N$ leads to the infinitesimal generator $(k_0 + K) \cal G$ where $K = \sum_{n=1}^N k_n$ measures how much of the $N$ available units of the resource is allocated to risky arms overall. 
This fact will play a crucial role in our analysis.

The generator ${\cal G}$ is that of a jump diffusion.
In the interior $\overset{\circ}{\Delta}_L$ of the simplex, its action on a $C^2$ function $u$ is given by
\begin{eqnarray*}
{\cal G}u(\pi) &\!=\!&
  \frac{1}{2\sigma^2}\,\sum_{i=1}^L\,\sum_{\ell=1}^L\,
  \pi_i\,\pi_\ell\,[\rho_i-\rho(\pi)][\rho_\ell-\rho(\pi)]\,
  \frac{\partial^2 u(\pi)}{\partial \pi_i\,\partial \pi_\ell}
\\ & & \mbox{} +
  \int_{\Rnotz} \left[u(j(\pi,h)) - u(\pi)\right]\,\nu(\pi)(dh)
\ -\
  \sum_{\ell=1}^L\,\pi_\ell
\left(\lambda_\ell - \lambda(\pi)\right)
\frac{\partial u(\pi)}{\partial \pi_\ell} \,,
\end{eqnarray*}
where
$$
  \rho(\pi) = \sum_{\ell=0}^L \pi_\ell\,\rho_\ell ,
\quad
  \nu(\pi) = \sum_{\ell=0}^L\pi_\ell\,\nu_\ell ,
\quad
  \lambda(\pi) = \sum_{\ell=0}^L \pi_\ell\,\lambda_\ell
$$
are the expected drift rate of the payoff-generating process, its expected L\'{e}vy measure, and the expected number of its jumps per unit of time, respectively, given the current belief $\pi$,
and
$$
j_\ell(\pi,h) = \frac{\pi_\ell\,\nu_\ell(dh)}{\nu(\pi)(dh)}
$$
is the revised probability of state $\ell$ after a lump-sum payoff of size $h$ arrives.
The first term captures the learning from the continuous part of the payoff-generating process;
the second term, the discrete belief revision upon the arrival of a lump-sum payoff;
and the third term, the gradual belief revision when no such lump-sum arrives.

For $L=1$, and hence $\pi = \pi_1$, we obtain the generator computed by Cohen and Solan (2013), with the first term simplifying to
$$
\frac{1}{2\sigma^2} (\rho_1 - \rho_0)^2 \pi^2 (1-\pi)^2 \, u''(\pi),
$$
the expression familiar from Bolton and Harris (1999, 2000).
It reflects the fact, established in Liptser and Shiryayev (1977, Theorem 9.1), that when there is no discontinuous payoff component ($\lambda_0 = \lambda_1 = 0$), then the posterior belief $\pi_t$ of a single agent who allocates his entire resource to the risky arm follows a diffusion process with zero drift and diffusion coefficient
$(\rho_1 - \rho_0)\,\sigma^{-1}\pi_t(1-\pi_t)$
relative to the agent's information filtration.\footnote{
More precisely, the belief evolves according to
$
  d\pi_t
= \sigma^{-1}\,\pi_t [\rho_1-\rho(\pi_t)] \,d\bar{Z}_t
$
where the {\it innovation\/} process $\bar{Z}_t$, given by
$
  d\bar{Z}_t
= \sigma^{-1} \left(\astrut{2} [\rho-\rho(\pi_t)]\,dt + \sigma\,dZ_t \right)
$,
is a Wiener process relative to the agent's information filtration.}
For $L > 1$, a generalization of Liptser and Shiryayev (1977, Theorem 9.1) shows that, from the agent's perspective, the corresponding belief process $\pi_t$ is a driftless $L$-dimensional diffusion with instantaneous variance-covariance matrix given by
$$
  \Cov\left[d\pi_{i,t}, d\pi_{\ell,t} \mid \pi_t \right]
  = \left[\pi_{i,t}   \,(\rho_i   - \rho(\pi_t))\,\sigma^{-1}\right]
    \left[\pi_{\ell,t}\,(\rho_\ell- \rho(\pi_t))\,\sigma^{-1}\right] dt,
$$
hence the structure of the first term in ${\cal G}u$.\footnote{
This generalization already appears in Veronesi (2000), for example.}

The second and third terms generalize their counterparts in Cohen and Solan (2013) to $L > 1$ in the obvious way.
In the special case that $L = 1$ and the size of lump-sum payoffs is uninformative (meaning that conditional on the arrival of a lump-sum, the distribution of its size does not depend on $\ell$), these terms reduce to
$$
\lambda(\pi) \left[ u\!\left( \frac{\pi \lambda_1}{(1-\pi) \lambda_0 + \pi \lambda_1}\right) - u(\pi) \right]
- (\lambda_1-\lambda_0) \pi (1-\pi) u'(\pi),
$$
as in Keller, Rady and Cripps (2005) and Keller and Rady (2010).

Note that we have not imposed any mutual absolute continuity assumptions on the measures $\nu_0,\ldots,\nu_L$.
As a consequence, lump-sum payoffs of a certain size may rule out certain states, so that the posterior belief jumps to a subsimplex of $\Delta_L$ of dimension lower than $L$.
Once this happens, Bayesian updating ensures that beliefs remain in this subsimplex.


\section{Symmetric Markov Perfect Equilibrium
\label{sec: equil}}

Suppose that all players except player $n$ use the strategy $\kappa^\dagger \in \cal K$,
and write $(\kappa_n,\kappa^\dagger_{\neg n})$ for the strategy profile that results when player $n$ uses the strategy $\kappa_n \in \cal K$.

When choosing $\kappa_n$, player $n$ faces a problem of optimal stochastic control of a jump diffusion, and $\kappa_n$ is a best response if and only if the payoff function $u_n(\cdot|\kappa_n,\kappa^\dagger_{\neg n})$ is the value function for that control problem.
The associated HJB equation is
\begin{equation} \label{eq: HJB-1}
  0 = \max_{k \in [0,1]} \left\{ \astrut{2} (1-k) s + k m(\pi) - f(\pi) + [k_0 + (N-1) \kappa^\dagger(\pi) + k] {\cal G} u(\pi) \right\}.
\end{equation}
Following Bolton and Harris (2000), we exploit the fact that $k_0 + (N-1) \kappa^\dagger(\pi) + k$ is positive (because of the background signal)
and rearrange the HJB equation as
$$
  0 = \max_{k \in [0,1]}
    \frac{s - f(\pi) + k [m(\pi) - s]}{k_0 + (N-1) \kappa^\dagger(\pi) + k}
  + {\cal G} u(\pi),
$$
which demonstrates that the set of maximizers does not depend on continuation values.
In a second step, we rewrite the HJB equation so that $k$ appears only in the denominator:
\begin{equation} \label{eq: HJB-2}
  0 = \max_{k \in [0,1]}
    \frac{[k_0 + (N-1) \kappa^\dagger(\pi)] [s - m(\pi)] - [f(\pi) - s]}{k_0 + (N-1) \kappa^\dagger(\pi) + k}
    - [s - m(\pi)] + {\cal G}u(\pi).
\end{equation}

Following Bolton and Harris (2000) again,
we define the \emph{incentive to experiment} by
$$
  I(\pi) = \frac{f(\pi) - s}{s - m(\pi)}
$$
when $m(\pi) < s$, and $\infty$ otherwise.
When $I(\pi) < k_0 + (N-1) \kappa^\dagger(\pi)$, the numerator in \eref{eq: HJB-2} is positive
and the maximum is achieved by $k=0$;
when $I(\pi) > k_0 + (N-1) \kappa^\dagger(\pi)$, the numerator is negative
and the maximum is achieved by $k=1$;
when $I(\pi) = k_0 + (N-1) \kappa^\dagger(\pi)$, the numerator is zero and the choice of $k$ is inconsequential.

There are three different ways, therefore, in which $k=\kappa^\dagger(\pi)$ can achieve the maximum in the HJB equation:
either $\kappa^\dagger(\pi) = 0$ and $I(\pi) \leq k_0$,
or $\kappa^\dagger(\pi) = 1$ and $I(\pi) \geq k_0 + N-1 $,
or $0 < \kappa^\dagger(\pi) < 1$ and $I(\pi) = k_0 + (N-1) \kappa^\dagger(\pi)$.
This pins down $\kappa^\dagger(\pi)$ in terms of the incentive to experiment, $I(\pi)$, the strength of the background signal, $k_0$, and the number of players, $N$:
\begin{equation}\label{eq: sym}
\kappa^\dagger(\pi) = \left\{ \begin{array}{ll}
0                           & \mbox{if } I(\pi) \leq k_0, \\
\frac{I(\pi) - k_0}{N-1}    & \mbox{if } k_0 < I(\pi) < k_0 + N - 1, \\
1                           & \mbox{if } I(\pi) \geq k_0 + N - 1.
\end{array}\right.
\end{equation}
As the partial derivatives of the incentive to experiment $I$ are clearly bounded on the compact set $\left\{ \pi \in \Delta_L: k_0 \leq I(\pi) \leq k_0 + N -1 \right\}$, the function $\kappa^\dagger$ is Lipschitz continuous and hence an element of $\cal K$.
Finally, it is straightforward to verify that $\kappa^\dagger$ is a reasonable strategy as defined in Section \ref{sec: model}.

\noindent \textbf{Proposition.}
\emph{All players using the strategy $\kappa^\dagger$ constitutes the unique symmetric Markov perfect equilibrium of the experimentation game.}

\noindent \proof{
Suppose that all players except player $n$ use the strategy $\kappa^\dagger$ defined in \eref{eq: sym}.
Let $u^*(\cdot|\kappa^\dagger_{\neg n})$ denote the value function of the control problem that player $n$ faces when choosing a best response,
and $u(\cdot|\kappa^\dagger,\kappa^\dagger_{\neg n})$ the player's payoff function when she also uses strategy $\kappa^\dagger$.
By definition, $u^*(\cdot|\kappa^\dagger_{\neg n}) \geq u(\cdot|\kappa^\dagger,\kappa^\dagger_{\neg n})$.

We show in the appendix that $u^*(\cdot|\kappa^\dagger_{\neg n})$ is a viscosity subsolution of the HJB equation \eref{eq: HJB-1} in the interior $\overset{\circ}{\Delta}_L$ of the $L$-dimensional simplex,
and $u(\cdot|\kappa^\dagger,\kappa^\dagger_{\neg n})$ a viscosity supersolution; cf.~Lemmas \ref{lem: subsolution}--\ref{lem: supersolution}.\footnote{
A definition of these concepts is also given in the appendix.}
As both functions vanish in all vertices of $\Delta_L$, the comparison principle for viscosity sub- and supersolutions established in Ishii and Yamada (1993, Theorem 3.1) allows us to conclude that $u^*(\cdot|\kappa^\dagger_{\neg n}) \leq u(\cdot|\kappa^\dagger,\kappa^\dagger_{\neg n})$; cf.~Corollary \ref{cor}.\footnote{
It is this comparison principle that implies the uniqueness result for viscosity solutions to the HJB equation alluded to in the introduction.}
As $u^*(\cdot|\kappa^\dagger_{\neg n}) = u(\cdot|\kappa^\dagger,\kappa^\dagger_{\neg n})$, therefore, all players using the strategy $\kappa^\dagger$ constitutes an equilibrium.\footnote{
The identity between these two functions further implies that the value function is both a sub- and a supersolution, confirming the claim made in the introduction that $u^*(\cdot|\kappa^\dagger_{\neg n})$ is a viscosity solution of the HJB equation.}
Uniqueness of this symmetric equilibrium follows from the arguments that led us from the HJB equation \eref{eq: HJB-1} to the representation \eref{eq: sym} for candidate equilibrium actions.
}

Figures~\ref{fig: simplex hi_s} and~\ref{fig: simplex lo_s} illustrate the case $L=2$.
(In both figures,
$\mu_0 = 2$, $\mu_1 = 5$, $\mu_2 = 8$, $N = 4$ and $k_0 = 0.2$;
$s = 6$ in Figure \ref{fig: simplex hi_s},
and $s = 4$ in Figure \ref{fig: simplex lo_s}.)
The solid lines are the boundaries of the sets of beliefs at which the equilibrium requires full experimentation ($\kappa^\dagger = 1$) and no experimentation ($\kappa^\dagger = 0$), respectively.
The dotted lines are level curves of $\kappa^\dagger$ for the experimentation intensities~0.2, 0.4, 0.6 and~0.8.
A comparison of the two figures exhibits the familiar property that a decrease in the reward from the safe arm gives the players an increased incentive to experiment.

\setlength {\unitlength} {0.65mm}  
\begin{figure}[h]
\begin{minipage}{.47\linewidth} \centering
  \setboolean{hi_s}{true}
  \input{UB-simplex.pic}
  \setboolean{hi_s}{false}
  \caption{
    Equilibrium actions for $L=2$}
and $\mu_0 < \mu_1 < s < \mu_2$  
  \label{fig: simplex hi_s}
\end{minipage}
\hfill
\begin{minipage}{.47\linewidth} \centering
  \setboolean{lo_s}{true}
  \input{UB-simplex.pic}
  \setboolean{lo_s}{false}
  \caption{
    Equilibrium actions for $L=2$}
and $\mu_0 < s < \mu_1 < \mu_2$
  \label{fig: simplex lo_s}
\end{minipage}
\end{figure}

Figures~\ref{fig: simplex hi_s} and~\ref{fig: simplex lo_s} further illustrate the fact that like the functions $m$, $f$, and $I$, the equilibrium strategy $\kappa^\dagger$ is non-decreasing in each component of $\pi = (\pi_1,\ldots,\pi_L)$;
thus, any shift in subjective probability mass from state 0 to another state weakly increases the intensity of experimentation.

Note that by equation \eref{eq: sym}, the set of beliefs for which $\kappa^\dagger(\pi) = 0$ is independent of the number of players.
This is a stark manifestation of the incentive to free-ride on information generated by others.
In the terminology coined by Bolton and Harris (1999), it means that there is no `encouragement effect':
the prospect of subsequent experimentation by other players provides a player \emph{no} incentive to increase the current intensity of experimentation and thereby shorten the time at which the information generated by the other players arrives.
Intuitively, this simply reflects our assumption that players do not discount future payoffs and hence are indifferent as to their timing.
Formally, the absence of the encouragement effect is a consequence of the linearity of the infinitesimal generator of posterior beliefs in $k_0 +K$:
as the value of future experimentation by other players is captured by a player's equilibrium continuation values,
yet best responses are independent of those continuation values,
there is no channel for future experimentation by others to impact current actions.

Free-riding can also be seen in the fact that $\kappa^\dagger$ is non-increasing in $N$, and decreasing where it assumes interior values.
Figure~\ref{fig: vary_N} illustrates this effect.
\setlength {\unitlength} {1mm}  
\begin{figure}[h] \centering
\begin{picture}(115.0,70.0)(-10.0,-10.0)
\put(  0,-4){\scalebox{0.33}{\includegraphics{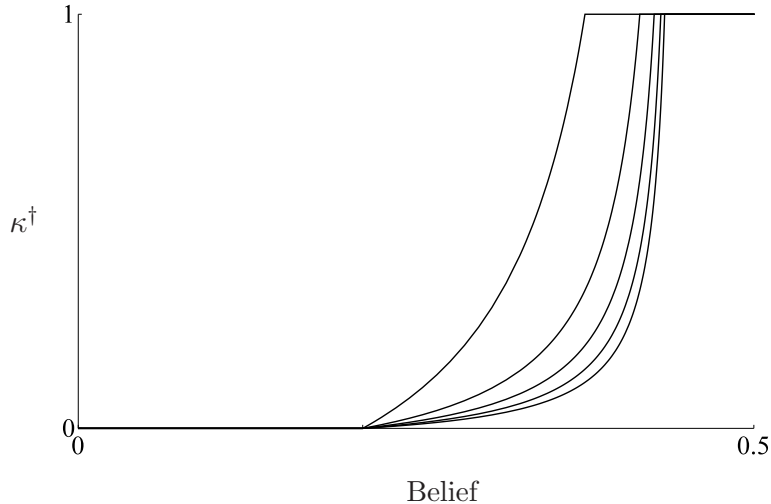}}}
\put( 50.0,-8.0){\makebox(0,0)[cc]{\small Belief}}
\put( -5.0,28.0){\makebox(0,0)[cc]{$\kappa^\dagger$}}
\end{picture}
  \caption{
    Equilibrium actions for $L=2$, $\pi_1=\pi_2$ and $N \in \{2,4,6,8,10\}$
    }
  \label{fig: vary_N}
\end{figure}
On the horizontal axis we set $\pi_1 = \pi_2$ and let that common belief range from~0 to~0.5:
so it is a slice through the simplex from the origin to the midpoint of the opposite edge.
(In this figure, the parameters are as in Figure \ref{fig: simplex hi_s} except that $N$ varies from~2 for the leftmost curve to~10 for the rightmost curve.)

The dependence of the overall intensity of experimentation on the number of players is less clear cut:
roughly speaking, $N\kappa^\dagger$ increases in $N$ at beliefs where $\kappa^\dagger$ requires exclusive use of the risky arm, but decreases at beliefs where both arms are used simultaneously.

Players also free-ride on the background information.
The less of it is available (i.e.\ the lower $k_0$), the more experimentation of their own the players perform; in particular, the set of beliefs at which they use the risky arm exclusively widens, and the set of beliefs at which they use the safe arm exclusively shrinks.
For $k_0 \downarrow 0$, in fact, the equilibrium strategy $\kappa^\dagger$ converges monotonically to the Markov strategy $\kappa^\dagger_0$ defined as follows:
$\kappa^\dagger_0(\pi) = 0$ when $f(\pi) = s$;
$\kappa^\dagger_0(\pi) = I(\pi)/(N-1)$ when $0 < I(\pi) < N-1$;
and $\kappa^\dagger_0(\pi) = 1$ when $I(\pi) \geq  N-1$.
This strategy obviously specifies the optimal action when $f(\pi) = s$ and, with $k_0 = 0$, the action $k = \kappa^\dagger_0(\pi)$ solves the maximization problem in equation \eref{eq: HJB-2} at all other beliefs.
Nevertheless, $\kappa^\dagger_0$ is not part of a symmetric MPE of the undiscounted game without background information.
The reason is that the strong long-run average criterion implies a payoff equal to $- \infty$ for the corresponding strategy profile in the interior of the simplex, so the HJB equation loses its meaning.\footnote{
This is most easily seen in the setting of Keller, Rady and Cripps (2005) with $L = 1$, no Brownian payoff component, no lump-sum payoffs in state 0, and a Poisson payoff process in state 1.
If state 0 is the true state, the belief $\pi_1$ converges deterministically to 0 from any prior in the open unit interval and, even though $\kappa^\dagger_0$ specifies a positive intensity of experimentation all along the way, this intensity decreases quite fast.
As a consequence, the convergence of the posterior to the truth is so slow in this state that the expected accumulated shortfall of received payoffs relative to the expected full-information payoff grows infinitely large.
Details are available from the authors upon request.}

It was already said in Section \ref{sec: model} that the presence of background information ensures exponentially fast convergence of beliefs to the degenerate distribution concentrated on the true state; this immediately implies that equilibrium actions converge exponentially fast to the full-information optimum as well.

As to the short-run dynamics of beliefs in equilibrium, the present framework permits the analysis of experimentation games in which large payoff increments are bad news, whereas smaller increments are good news.\footnote{
In Keller, Rady and Cripps (2005) and Keller and Rady (2010, 2015) lump-sum sizes are completely uninformative, while in Cohen and Solan (2013) lump-sums are informative, but always good news.}
For example, let $L = 1$ for simplicity, with
$\rho_0 = \rho_1$
and
$\lambda_0 = \lambda_1$.
Assume that the payoff increments are in the set
$\{-10, -5, 5, 10\}$.
For the `good' arm, the associated probabilities of a lump-sum of that size are
$(0.1, 0.3, 0.5, 0.1)$,
so the expected increment is $1$;
for the `bad' arm, the probabilities are
$(0.5, 0.1, 0.1, 0.3)$,
and the expected increment is $-2$.
When a  payoff increment occurs,
the belief jumps --
up if the increment is moderate
($-5$ and $5$ are relatively more likely if the arm is `good'),
and down if the increment is extreme
($-10$ and $10$ are relatively more likely if the arm is `bad').
So, in this stripped-down illustration, an arrival of the largest possible payoff increment is bad news, and may well cause the players to stop experimenting.

\section{Continuous State Spaces and Sufficient Statistics}
\label{sec: examples}

This section presents two specifications of priors and payoff-generating processes that fall outside the framework of Section \ref{sec: model} but still permit the same analysis as in Sections \ref{sec: gen} and \ref{sec: equil}.
In both settings, the unknown state of the world is drawn from a continuous distribution of unbounded support, with conjugate priors ensuring that the players' information is captured by a two-dimensional sufficient statistic, which can serve as the state variable for Markov strategies.\footnote{
The unbounded state space requires adjustments to the proof (via uniqueness of viscosity solutions to the HJB equation) that every player using the strategy $\kappa^\dagger$ constitutes an MPE of the game;
we omit the details here.}
Models in which agents have beliefs and observe stochastic processes like those in Sections~\ref{example: BM,normal} and \ref{example: PP,gamma} can be found in Jovanovic (1979) and Moscarini and Squintani (2010), respectively.


\subsection{Brownian Payoffs, Normal Prior
\label{example: BM,normal}}

Suppose that the payoff-generating processes and the background signal are of the form
$$
X^n_t = \mu\,t+\sigma Z^n_t,
$$
where the $Z^n$ are independent standard Wiener processes and nature draws the unknown drift $\mu$ from a normal distribution with mean $m_0$ and precision $\tau_0 > 0$.
This is also the players' common prior.
Given the Gaussian processes they observe, players then believe at time $t$ that $\mu$ is distributed according to a normal distribution with some mean $m_t$ and precision $\tau_t > 0$;
see DeGroot (1970, Chapter~9), for example.
The pair $\pi_t=(m_t,\tau_t)$ constitutes a sufficient statistic for the updating of beliefs, therefore.
Given a generic $\pi=(m, \tau) \in \R \, \times \, ]0,\infty[\,$,
the corresponding probability density function for $\mu$ is
$g(\mu; \pi) = \tau^{1/2}\phi\left((\mu-m)\tau^{1/2}\right)$,
where $\phi$ denotes the standard normal density.
Let $G(\cdot; \pi)$ denote the associated cumulative distribution function.

As in Section \ref{sec: gen}, consider a single player allocating his entire resource to the risky arm.
Following Chernoff (1968, Lemma 4.1) or Liptser and Shiryayev (1977, Theorem 10.1),
$\tau_t$ increases deterministically at the rate $\sigma^{-2}$
and $m_t$ is a driftless diffusion process with diffusion coefficient $\sigma^{-1}\,\tau_t^{-1}$
relative to the player's information filtration.\footnote{
More precisely, it can be shown that
$
  dm_t = \sigma^{-1}\,\tau_t^{-1}\,d\bar{Z}_t
\mbox{ and }
  d\tau_t = \sigma^{-2}\,dt
$
where, now, the innovation process is
$d\bar{Z}_t
= \sigma^{-1} \left(\astrut{2} [\mu-m_t]\,dt + \sigma\,dZ_t \right)
$.
Note that the expression equivalent to that for $dm_t$ to be found in equation~(9) of Jovanovic (1979) omits the term $[\mu-m_t]\,dt$.}
As a result, we see that
$$
  {\cal G}u(\pi) =
  \frac{1}{\sigma^2} \left[ \frac{1}{2 \tau^2}\,\frac{\partial^2 u(\pi)}{\partial m^2}
                   + \frac{\partial u(\pi)}{\partial\tau} \right]
$$
for any function of class $C^{2,1}$.
By the same arguments as in Section \ref{sec: gen}, moreover, the generator associated with time-invariant intensities $(k_0,k_1,\ldots,k_N) \in [0,1]^{N+1}$ is again $(k_0 + K) \cal G$.

Since the precision $\tau_t$ increases over time, the relevant state space is the half-plane $\Pi = \R \, \times \, [\tau_0,\infty[\,$.
As to admissible strategies, we take ${\cal K}$ to be the set of all functions $\kappa\!: \Pi \rightarrow [0,1]$ such that $\kappa \tau^{-1}$ is Lipschitz continuous on $\Pi$.
Given a strategy profile $(\kappa_1,\ldots,\kappa_N) \in {\cal K}^N$, the sum $K = \sum_{n=1}^N \kappa_n$ also lies in $\cal K$, and the system we need to solve is
$$
  dm = K(m,\tau)\, \tau^{-1} \sigma^{-1} d\bar{Z} ,
\quad
  d\tau = K(m,\tau)\, \sigma^{-2} dt .
$$
The change of variable $\eta = \ln \tau$ transforms this into
$dm = K(m,e^\eta)\,e^{-\eta} \sigma^{-1} d\bar{Z}$
and
$d\eta = K(m,e^\eta)\,e^{-\eta} \sigma^{-2} dt$;
as $K(m,e^\eta)\,e^{-\eta}$ is Lipschitz continuous in $(m,\eta)$ on $\R \, \times \, [\ln\tau_0,\infty[\,$, this system has a unique solution, as was to be shown.

We can now replicate the arguments of Section \ref{sec: equil} in the present setting.
As a first step, we compute the expected current payoff $m(\pi)$, the expected full-information payoff $f(\pi)$, and the incentive to experiment $I(\pi)$.
The expected current payoff $m(\pi)$ is simply the projection of $\pi$ on its first component.
For the expected full-information payoff, we have
$$
  f(\pi) = s\,\Phi(z) + m\,[1 - \Phi(z)] + \tau^{-1/2}\phi(z),
$$
where $z = (s-m) \tau^{1/2}$
and $\Phi$ denotes the standard normal cumulative distribution function.
To see this, note first that
$f(\pi) = s G(s; \pi) + \int_s^\infty \mu\, g(\mu ; \pi) \, d\mu$.
We trivially obtain
$G(s; \pi) = \int_{-\infty}^s g(\mu; \pi) \, d\mu = \int_{-\infty}^{z} \phi(x) \, dx = \Phi(z)$.
Since $g(\mu; \pi) \propto \exp\left(-\divn{1}{2}(\mu-m)^2 \tau\right)$, moreover, we have
$dg(\mu; \pi) = -(\mu-m)\tau\,g(\mu; \pi)\,d\mu$
and so
$\mu\,g(\mu; \pi)\,d\mu = m\,g(\mu; \pi)\,d\mu - \tau^{-1}\,dg(\mu; \pi)$,
implying
\begin{eqnarray*}
  \int_s^\infty \mu \, g(\mu ; \pi) \, d\mu
&=& \int_s^\infty m\,g(\mu; \pi)\,d\mu - \int_s^\infty \tau^{-1}\,dG(\mu; \pi) \\
&=& m\,[1 - G(s; \pi)] + \tau^{-1}\,g(s; \pi)
\ = \ m\,[1 - \Phi(z)] + \tau^{-1/2}\,\phi(z).
\end{eqnarray*}

The above representation makes it straightforward to verify that $f$ is strictly increasing in $m$ and strictly decreasing in $\tau$.\footnote{
Alternatively,
since $\max\{s, \mu\}$ is increasing in $\mu$,
a first-order stochastic dominance argument can be used to establish that
$\partial f(\pi)/\partial m > 0$,
and
since $\max\{s, \mu\}$ is convex in $\mu$,
a second-order stochastic dominance argument can be used to establish that
$\partial f(\pi)/\partial\tau < 0$.}
This implies that $I$ and $\kappa^\dagger$ as defined in \eref{eq: sym} are non-decreasing in $m$ and non-increasing in $\tau$.

When $m < s$, we have
$$
I(\pi)
= \frac{s\,\Phi(z) + m\,[1 - \Phi(z)] + \tau^{-1/2}\phi(z) - s}{s-m}
= \Phi(z) - 1 + z^{-1} \phi(z).
$$
In the appendix, we verify that $\kappa^\dagger \in \cal K$ by showing that $I\tau^{-1}$ is Lip\-schitz continuous on
the set $\left\{ \pi \in \Pi: k_0 \leq I(\pi) \leq k_0 + N -1 \right\}$.
This is more involved than in scenarios with a discrete prior because the set in question is unbounded.

Figure~\ref{fig: BM,normal} illustrates equilibrium actions as a function of the posterior mean $m$ and variance $\tau^{-1}$.
(In this figure,
$s = 6$ and $N = 4$.)
As in Figures~\ref{fig: simplex hi_s}--\ref{fig: simplex lo_s}, the solid curves are the boundaries of the sets of beliefs at which the equilibrium requires full experimentation or no experimentation, and the dashed lines are level curves for $\kappa^\dagger$ equal to~0.2, 0.4, 0.6 and~0.8.
All these curves are downward sloping; as one would expect, there is a trade-off between mean and variance with the latter capturing the `option value' of experimentation.
In particular, a very high variance is needed to induce a high intensity of experimentation at low means.
As the mean approaches the safe flow payoff, the level curves become steeper and steeper so that the posterior variance has a diminishing impact on the intensity with which the players explore the risky arm.
\setlength {\unitlength} {1mm}  
\begin{figure}[t] \centering
  \begin{picture}(115.0,70.0)(-10.0,-10.0)
    \put(  0,-4){\scalebox{0.33}{\includegraphics{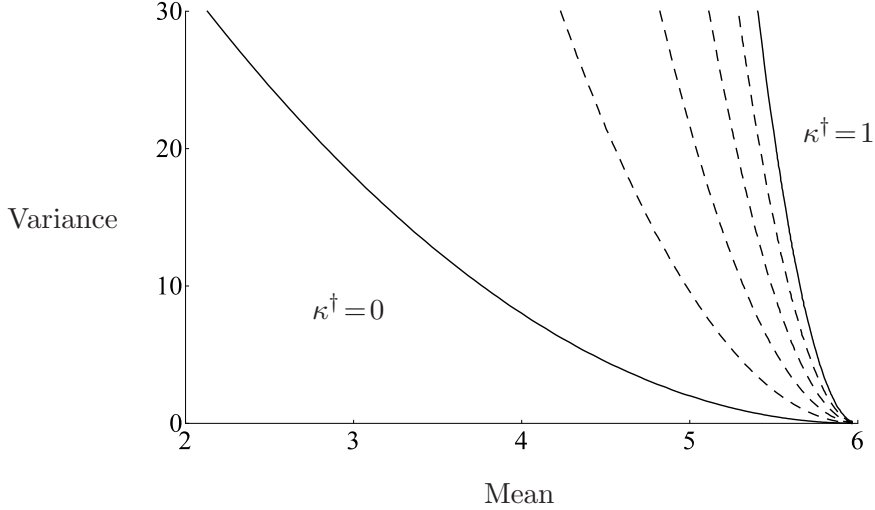}}}
    \put( 47.75,-9.0){\makebox(0,0)[cc]{\small Mean}}
    \put( -5.0,27.5){\makebox(0,0)[rc]{\small Variance}}
    \put( 30.0,17.5){\makebox(0,0)[tr]{\small $\kappa^\dagger\!=\!0$}}
    \put( 85.0,37.5){\makebox(0,0)[bl]{\small $\kappa^\dagger\!=\!1$}}
  \end{picture}
  \caption{
    Equilibrium actions for Brownian payoffs and normal prior
  }
  \label{fig: BM,normal}
\end{figure}


\subsection{Poisson Payoffs, Gamma Prior
\label{example: PP,gamma}}

Let $s > 0$ for the safe arm.
Suppose that the payoff-generating processes and the background signal are independent Poisson processes whose unknown common intensity $\mu$
is drawn from a gamma distribution with parameters $\alpha_0 > 0$ and $\beta_0 > 0$.
This is also the players' common prior.
Given the processes they observe, players then believe at time $t$ that $\mu$ is distributed according to a gamma distribution with some parameters $\alpha_t > 0$ and $\beta_t > 0$, which together constitute a sufficient statistic again;
see DeGroot (1970, Chapter~9), for example.
Given a generic $\pi=(\alpha, \beta) \in \ ]0,\infty[^2$, the probability density function for $\mu$ is
$
  g(\mu ; \pi)
= [{\beta^\alpha}/{\Gamma(\alpha)}] \mu^{\alpha-1} e^{-\beta\mu};
$
the mean and variance of $\mu$ are $\alpha/\beta$ and $\alpha/\beta^2$, respectively.
We again write $G(\cdot; \pi)$ for the corresponding cumulative distribution function.

Once more, consider a single player allocating his entire resource to the risky arm.
He expects to obtain a positive increment between $t$ and $t+dt$ with probability $(\alpha_t/\beta_t)\,dt$,
in which case Bayes' rule implies that $\pi_t$ jumps to $(\alpha_t+1,\beta_t)$;
with probability $1 - (\alpha_t/\beta_t)\,dt$, there is no such increment and
$
  d\pi_t = (d\alpha_t,d\beta_t) = (0,dt).
$
Thus,
$\alpha$ counts arrivals of increments
and
$\beta$ measures the time that has elapsed --
again, see DeGroot (1970, Chapter~9).
As a consequence, we have\footnote{
Up to a change of variables from $(\alpha,\beta)$ to $(\beta,m)$ with $m = \alpha/\beta$, this generator also appears in Ding and Ryzhov (2016).}
$$
  {\cal G}u(\pi) =
  \frac{\alpha}{\beta} \left[u(\alpha+1, \beta) - u(\pi)\right]
+ \frac{\partial u(\pi)}{\partial\beta} \,.
$$
Once more, the generator associated with time-invariant intensities $(k_0,k_1,\ldots,k_N) \in [0,1]^{N+1}$ is $(k_0 + K) \cal G$.

Given that $\alpha_t$ and $\beta_t$ increase over time, and $\alpha_t$ can only do so in unit increments, the relevant state space is
$\Pi = \{ \alpha_0 + j: j = 0,1,2,\ldots \} \times [\beta_0,\infty[\,$.
For $\cal K$, we choose the set of all functions $\kappa: \Pi \rightarrow [0,1]$ such that $\kappa(\alpha_0 + j,\cdot)$ is right-continuous and piecewise Lipschitz continuous for all $j$.
Starting from any $\pi \in \Pi$, any strategy profile $(\kappa_1,\ldots,\kappa_N) \in {\cal K}^N$ induces a well-defined law of motion for $\pi_t$.

As the unknown intensity $\mu$ is also the risky arm's average payoff per unit of time,
we see that the expected current payoff is $m(\pi) = \alpha/\beta$.
The expected full-information payoff is
$$
  f(\pi) = s\,G(s; \pi)
         + \frac{\alpha}{\beta}\,[1 - G(s; \alpha+1, \beta)],
$$
with the second term obtained as follows:
\begin{eqnarray*}
  \int_s^\infty \mu \, g(\mu ; \pi) \,d\mu
&=& \int_s^\infty \mu \, \frac{\beta^\alpha}{\Gamma(\alpha)}\,\mu^{\alpha-1} e^{-\beta\mu} \, d\mu
 = \frac{\alpha}{\beta} \int_s^\infty \frac{\beta^{\alpha+1}}{\alpha \Gamma(\alpha)}\,\mu^{\alpha} e^{-\beta\mu} \, d\mu
\\
&=&  \astrut{4.5} \frac{\alpha}{\beta} \int_s^\infty \frac{\beta^{\alpha+1}}{\Gamma(\alpha+1)}\,\mu^{\alpha} e^{-\beta\mu} \, d\mu
= \frac{\alpha}{\beta} \int_s^\infty g(\mu; \alpha+1, \beta) \, d\mu
\\
&=&  \astrut{4.5} \frac{\alpha}{\beta}\,[1 - G(s; \alpha+1, \beta)].
\end{eqnarray*}

The formula for $f$ makes it straightforward to verify that, exactly like $m$, this function is strictly increasing in $\alpha$ and strictly decreasing in $\beta$.\footnote{
Alternatively, for $\alpha' > \alpha''$
the likelihood ratio $g(\alpha; \alpha', \beta) / g(\alpha; \alpha'', \beta)$
is increasing,
and
for $\beta' > \beta''$
the likelihood ratio $g(\alpha; \alpha, \beta') / g(\alpha; \alpha, \beta'')$
is decreasing.
Since the likelihood-ratio ordering implies first-order stochastic dominance,
$f$ has the stated monotonicity properties.}
Consequently, the incentive to experiment $I$ and the strategy $\kappa^\dagger$ as defined in \eref{eq: sym} are non-decreasing in $\alpha$ and non-increasing in $\beta$.

For $m(\pi) < s$, we have
$$
I(\pi)
= \frac{s\,G(s; \alpha, \beta) + \divn{\alpha}{\beta}\,[1 - G(s; \alpha+1, \beta)] - s}{s-\divn{\alpha}{\beta}}
= \frac{s\,G(s; \alpha, \beta) - \divn{\alpha}{\beta}\,G(s; \alpha+1, \beta)}{s-\divn{\alpha}{\beta}} - 1.
$$
In the appendix, we verify that $\kappa^\dagger \in \cal K$ by showing for any fixed $\alpha$ that $I(\alpha,\cdot)$ has a bounded first derivative
when $m(\pi)<s$.

Figure~\ref{fig: PP,gamma} illustrates the mean-variance trade-off in equilibrium actions for Poisson payoffs and gamma prior.
(Here,
as in the example with Brownian payoffs and normal prior,
$s = 6$ and $N = 4$;
the curves shown are thus the exact counterparts of those in Figure~\ref{fig: BM,normal}.)
To compute the level curves, one uses the fact that the shape parameter $\alpha$ equals the squared mean of the gamma distribution divided by its variance,
and $\beta$ is $\alpha$ divided by the mean.
The similarity to Figure~\ref{fig: BM,normal} is striking;
a closer comparison reveals that the level curves in the Brownian-normal case are somewhat steeper than those in the Poisson-gamma case.
This is because in the former, an increase in the variance induces a mean-preserving spread for the random variable $\alpha$ on the whole real axis, whereas in the latter, the mean-preserving spread is concentrated on the positive half-axis and thus raises the option value of experimentation by more.
\setlength {\unitlength} {1mm}  
\begin{figure}[t] \centering
  \begin{picture}(115.0,70.0)(-10.0,-10.0)
    \put(  0,-4){\scalebox{0.33}{\includegraphics{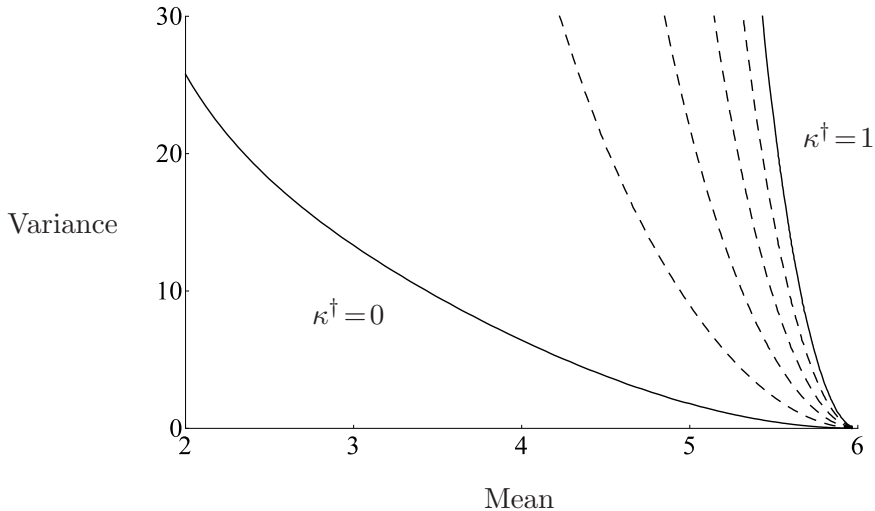}}}
    \put( 47.75,-9.0){\makebox(0,0)[cc]{\small Mean}}
    \put( -5.0,27.5){\makebox(0,0)[rc]{\small Variance}}
    \put( 30.0,17.5){\makebox(0,0)[tr]{\small $\kappa^\dagger\!=\!0$}}
    \put( 85.0,37.5){\makebox(0,0)[bl]{\small $\kappa^\dagger\!=\!1$}}
  \end{picture}
  \caption{
    Equilibrium actions for Poisson payoffs and gamma prior
  }
  \label{fig: PP,gamma}
\vspace{1.5ex}
\end{figure}


\section{Concluding Remarks}
\label{sec: concl}

We have seen that when rewards from the risky arm are generated by IID L\'{e}vy processes with an unknown average payoff per unit of time, the players' strategy in the symmetric MPE of the undiscounted experimentation game depends only -- and in a very simple functional form -- on the safe payoff, the expected current payoff of the risky arm, and the expected full-information payoff.
Given a finite set from which nature draws the unknown average payoff, the equilibrium strategy is then independent of the actual specification of the payoff-generating processes.

As to the settings with a continuous prior,
recall that in the Brownian-normal case the precision of the posterior distribution increases unboundedly with time,
as does the inverse of the variance in the Poisson-gamma case.
Consequently, the posterior probability density function becomes concentrated on a narrow domain of the support.
If we approximated the normal or gamma distribution with a discrete distribution then, over time, the beliefs would become more and more concentrated on the discrete values closest to the true parameter --
this suggests that we could take the `engineering' approach and focus on discrete distributions, with the specification of the payoff-generating processes being irrelevant.\footnote{
But note that if the two closest neighbours of the true average payoff $\mu$ per unit of time are $\mu_\ell$ and $\mu_{\ell+1}$ with $\mu_\ell < \mu < \mu_{\ell+1}$,
then, although $m(\pi_T) \simeq \mu$ for large $T$,
we would have $\Var[\mu|\pi_T] \simeq (\mu_{\ell+1}-\mu)(\mu-\mu_\ell)$, which is bounded away from zero.}

Letting the discount rate go to zero is going to make the analysis easier in many dynamic settings, but it remains unclear, in general, whether the simplification will be as great as in the present case.
Candidates for optimal strategies or best responses may be easier to identify in the undiscounted limit, but there remains the need to obtain a well-defined law of motion, which may again require restrictions such as Lipschitz continuity and could even lead to existence problems.
Nevertheless, we believe that the strong long-run average criterion has the potential to prove useful in other contexts, especially since strategies which are optimal under this criterion will shed light on (at least approximately) optimal behaviour for small positive discount rates.

\AppendixIn                
\addtocounter{section}{1}
\setcounter{equation}{0}

\subsubsection*{Boundedness of Payoffs from Reasonable Strategies}

We present the case $L = 1$ only, so that $\ell \in \{0,1\}$, $\pi = \pi_1 \in [0,1]$, $\mu_0 < s < \mu_1$, $m(\pi) = (1-\pi) \mu_0 + \pi \mu_1$ and $f(\pi) = (1-\pi) s + \pi \mu_1$.
Suppose first that the L\'{e}vy measures $\nu_0$ and $\nu_1$ are non-trivial and equivalent.

For the description of the evolution of beliefs, it is convenient to work with the log odds ratio
$$
\omega_t = \ln \frac {\pi_t}{1-\pi_t}\,,
$$
so that
$$
\pi_t = \frac{e^{\omega_t}}{1+e^{\omega_t}} \quad \text{and} \quad 1-\pi_t = \frac{e^{- \omega_t}}{1+e^{- \omega_t}}\,.
$$
\p
\begin{lemma} \label{lem: bound}
There exists a constant $C > 0$ such that for all $x,y \in \R$,
$$
\frac{e^{x+y}}{1+e^{x+y}} \leq \frac{e^x}{1+e^x} + \frac{e^x}{(1+e^x)^2} \, y + C \, \frac{e^x}{(1+e^x)^3} \, y^2.
$$
\end{lemma}
\proof{
For
$$
f(x,y) = \frac{e^{x+y}}{1+e^{x+y}}\,,
$$
we compute the partial derivatives
$$
f_y(x,y) = \frac{e^{x+y}}{(1+e^{x+y})^2}\,,
\quad
f_{yy}(x,y) = \frac{e^{x+y}(1-e^{x+y})}{(1+e^{x+y})^3}\,.
$$
For fixed $x$, the function $f(x,\cdot)$ thus has the following second-order Taylor approximation around $y_0 = 0$:
$$
f(x,y) \approx \frac{e^x}{1+e^x} + \frac{e^x}{(1+e^x)^2} \, y + \frac{1}{2} \frac{e^x(1-e^x)}{(1+e^x)^3} \, y^2.
$$
As $1-e^x \leq 1$, we have the local (with respect to the second variable) upper bound
$$
f(x,y) \leq \frac{e^x}{1+e^x} + \frac{e^x}{(1+e^x)^2} \, y + \frac{1}{2} \frac{e^x}{(1+e^x)^3} \, y^2.
$$
Replacing the factor $\half$ in the last term by a sufficiently large constant $C$ ensures a global upper bound.\footnote{
Numerical computations suggest that $C=2$ is large enough.}
}

Suppose now that starting from $\pi_0 = \pi$ (and corresponding $\omega_0=\omega$), the players use the strategy profile $(\kappa_1,\ldots,\kappa_N) \in {\cal K}^N$.
By an extension of the results in Cohen and Solan (2013, Section 3.2) to more than one agent, the log odds ratio at time $t>0$ can be written as
$$
\omega_t = \omega + \eta_\ell \left[ k_0 t + \sum_{n=1}^N \int_{0}^{t}\kappa_n(\pi_{s-})\,ds \right] + M^\ell_t,
$$
where
$$
\eta_\ell = (-1)^{\ell+1} \frac{(\rho_1 - \rho_0)^2}{2 \sigma^2} - (\lambda_1 - \lambda_0) + \int_{\Rnotz} \ln \frac{\nu_1}{\nu_0}(h) \, \nu_\ell(dh),
$$
$\frac{\nu_1}{\nu_0}$ is the Radon-Nikodym derivative of $\nu_1$ with respect to $\nu_0$,
and $M^\ell$ is a martingale under the probability measure $\Pr_\ell$ associated with state $\ell$.
The expectation and variance of $M^\ell$ under this measure, moreover, satisfy
$\Exp_\ell[M^\ell_t]= 0$ and $\Var_\ell[M^\ell_t] \leq C_\ell \, t$
for all $t$ and a positive constant $C_\ell$.\footnote{
For any fixed action profile, $M^\ell$ has stationary increments, so its variance grows linearly with time.
$C_\ell$ can be chosen as the rate at which the variance grows when all players use the risky arm exclusively.}

As $\ln x < x - 1$ for all positive $x \neq 1$, and $\frac{\nu_1}{\nu_0} = (\frac{\nu_0}{\nu_1})^{-1}$,
one sees that
$$
\int_{\Rnotz} \ln \frac{\nu_1}{\nu_0}(h) \, \nu_0(dh) < \lambda_1 - \lambda_0 < \int_{\Rnotz} \ln \frac{\nu_1}{\nu_0}(h) \, \nu_1(dh)
$$
unless $\nu_1 = \nu_0$, in which case the inequality $\mu_1 > \mu_0$ implies $\rho _1 > \rho_0$.
So $\eta_0 < 0 < \eta _1$.
As $\kappa_n \geq 0$ for all $n$, this in turn implies
$$
\omega + \eta_0 k_0 t + M^0_t \geq \omega_t \geq \omega + \eta_1 k_0 t + M^1_t.
$$
By Lemma \ref{lem: bound},
$$
\pi_t
\leq \frac{e^{\omega + \eta_0 k_0 t + M^0_t}}{1+e^{\omega + \eta_0 k_0 t + M^0_t}}
\leq \frac{e^{\omega + \eta_0 k_0 t}}{1+e^{\omega + \eta_0 k_0 t}}
+ \frac{e^{\omega + \eta_0 k_0 t}}{(1+e^{\omega + \eta_0 k_0 t})^2} \, M^0_t
+ C \, \frac{e^{\omega + \eta_0 k_0 t}}{(1+e^{\omega + \eta_0 k_0 t})^3} \, (M^0_t)^2
$$
and
$$
1-\pi_t
\leq \frac{e^{- \omega - \eta_1 k_0 t - M^1_t}}{1+e^{- \omega - \eta_1 k_0 t - M^1_t}}
\leq \frac{e^{- \omega - \eta_1 k_0 t}}{1+e^{- \omega - \eta_1 k_0 t}}
- \frac{e^{- \omega - \eta_1 k_0 t}}{(1+e^{- \omega - \eta_1 k_0 t})^2} \, M^1_t
+ C \, \frac{e^{- \omega - \eta_1 k_0 t}}{(1+e^{- \omega - \eta_1 k_0 t})^3} \, (M^1_t)^2.
$$
Writing $C'_\ell = C C_\ell$, we thus have
$$
\Exp_0[\pi_t]
\leq e^{\omega + \eta_0 k_0 t} \left( 1 + C \Var_0[M^0_t] \right)
= e^{\omega + \eta_0 k_0 t} (1 + C_0' t)
= \frac{\pi}{1-\pi} \, e^{\eta_0 k_0 t} (1 + C_0' t)
$$
and
$$
\Exp_1[1-\pi_t]
\leq e^{-\omega - \eta_1 k_0 t} \left( 1 + C \Var_1[M^1_t] \right)
= e^{-\omega - \eta_1 k_0 t} (1 + C_1' t)
= \frac{1-\pi}{\pi} e^{- \eta_1 k_0 t} (1 + C_1' t).
$$

Now let player $n$ use a reasonable strategy.
Then there is a constant $C_2 > 0$ such that
$$
[1-\kappa_n(\pi)] s + \kappa_n(\pi) m(\pi) - f(\pi) \geq C_2 \left[ \max\{s,m(\pi)\} - f(\pi) \right]
$$
for all $\pi$.\footnote{
This is because $[1-\kappa_n(\pi)] s + \kappa_n(\pi) m(\pi) = \max\{s,m(\pi)\}$ in neighbourhoods of $\pi = 0$ and 1,
and $\max\{s,m(\pi)\} - f(\pi)$ is bounded away from zero outside these neighbourhoods.}
Note that
$\max\{s,m(\pi)\} - f(\pi)$
is bounded below by
$s - f(\pi) = \pi \, (s-\mu_1)$
and by
$m(\pi) - f(\pi) = (1-\pi) \, (\mu_0-s)$.

Given the prior belief $\pi_0 = \pi$, the player uses the expectation operator $\Exp_\pi = (1-\pi) \Exp_0 + \pi \Exp_1$ to compute her objective function.
Thus,
\vspace{-1ex}
\begin{eqnarray*}
\lefteqn{u_n(\pi|\kappa_1,\ldots,\kappa_N)} \\
& \geq & (1-\pi) C_2 (s-\mu_1) \Exp_0\!\! \left[ \int_0^\infty  \pi_t \,dt \right] + \pi C_2 (\mu_0-s) \Exp_1\!\! \left[ \int_0^\infty (1-\pi_t) \,dt \right] \\
& = & \astrut{4} (1-\pi) C_2 (s-\mu_1) \int_0^\infty \Exp_0[\pi_t] \,dt + \pi C_2 (\mu_0-s) \int_0^\infty \Exp_1[1-\pi_t] \,dt \\
& \geq & \astrut{4} \pi C_2 (s-\mu_1) \int_0^\infty e^{\eta_0 k_0 t} (1 + C_0' t) \,dt + (1-\pi) C_2 (\mu_0-s) \int_0^\infty  e^{-\eta_1 k_0 t} (1 + C_1' t) \,dt \\
& = & \astrut{4} \pi C_2 (s-\mu_1) \frac{C_0' - \eta_0 k_0}{\eta_0^2 k_0^2} + (1-\pi) C_2 (\mu_0-s) \frac{C_1' + \eta_1 k_0}{\eta_1^2 k_0^2}\,.
\end{eqnarray*}
This is the desired result.

Next, suppose that the L\'{e}vy measure $\nu_1$, say, is not absolutely continuous with respect to $\nu_0$.
Take a $\nu_0$-null set $B \subseteq \Rnotz$ with $\nu_1(B) > 0$.
In state $\ell=1$, we then have
$\Pr_1[\pi_t = 1] \geq 1 - e^{-\nu_1(B) t}$,
so that
$$
\Exp_1[1-\pi_t]
= \Pr_1[\pi_t = 1] \cdot 0 + \Pr_1[\pi_t < 1] \cdot \Exp_1[1-\pi_t | \pi_t < 1]
\leq \Pr_1[\pi_t < 1]
\leq e^{-\nu_1(B) t}.
$$
This exponential convergence again allows us to compute an upper bound for $ \int_0^\infty \Exp_1[1-\pi_t] \,dt$.

Finally, if both L\'{e}vy measures are trivial, the inequality $\eta_0 < 0 < \eta _1$ holds trivially, and the result follows as above.

\subsubsection*{Viscosity Solutions of the HJB Equation}

Consider a nonempty, open, connected and bounded set $\Omega \subset \R^L$.
Denote the set of all symmetric $L \times L$ matrices by $\S^L$.
Let $H \in C(\Omega \times \R^L \times \S^L \times \R)$ satisfy
$$
H(x,p,X+Y,d) \geq H(x,p,X,d+q)
$$
for all $(x,p,X,d) \in \Omega \times \R^L \times \S^L \times \R$, all positive semidefinite $Y \in \S^N$ and all $q \geq 0$.\footnote{
Note that the variables $X$ and $Y$ just introduced are unrelated to the objects for which we use these symbols in the main text.}

We are interested in solutions $u\!: \overline{\Omega} \to \R$ of boundary value problems of the form
\begin{eqnarray}
  H(x,Du,D^2u,u-Mu) = 0 & & \text{in } \Omega, \label{eq: interior} \\
  u = v  & & \text{on } \partial\Omega, \label{eq: boundary}
\end{eqnarray}
where $Du$ and $D^2u$ are the gradient and the Hessian matrix of $u$, respectively,
$M$ is an operator mapping $C(\overline{\Omega})$ into itself,
and $v \in C(\overline{\Omega})$.

A function $u \in C(\overline{\Omega})$ is called a \emph{viscosity subsolution} of \eref{eq: interior}
if for every $\phi \in C^2(\overline \Omega)$ and every $x_0 \in \Omega$ such that $\phi \geq u$ on $\overline \Omega$ and $\phi(x_0) = u(x_0)$,
$$
H(x_0,D\phi(x_0),D^2\phi(x_0),\phi(x_0)-M\phi(x_0)) \geq 0.
$$
Analogously,
a function $u \in C(\overline{\Omega})$ is called a \emph{viscosity supersolution} of \eref{eq: interior}
if for every $\phi \in C^2(\overline \Omega)$ and every $x_0 \in \Omega$ such that $\phi \leq u$ on $\overline \Omega$ and $\phi(x_0) = u(x_0)$,
$$
H(x_0,D\phi(x_0),D^2\phi(x_0),\phi(x_0)-M\phi(x_0)) \leq 0.
$$
Finally, $u \in C(\overline{\Omega})$ is called a \emph{viscosity solution} of \eref{eq: interior} if it is a viscosity sub- and supersolution of \eref{eq: interior}.

The HJB equation \eref{eq: HJB-1} and its reformulation \eref{eq: HJB-2} are both of the form \eref{eq: interior} with $\Omega = \overset{\circ}{\Delta}_L$, the operator in question being
$$
Mu(\pi) = \frac{1}{\lambda(\pi)} \int_{\Rnotz} u(j(\pi,h)) \,\nu(\pi)(dh).
$$
By the arguments that led us from \eref{eq: HJB-1} to \eref{eq: HJB-2} in Section \ref{sec: equil}, these equations have the same viscosity solutions.
We will refer to either equation as the HJB equation in what follows.

Suppose that all players except player $n$ use the strategy $\kappa^\dagger$ defined in \eref{eq: sym}.
Let $u^*(\cdot|\kappa^\dagger_{\neg n})$ denote the value function of the control problem that player $n$ faces when choosing a best response,
and $u(\cdot|\kappa^\dagger,\kappa^\dagger_{\neg n})$ the player's payoff function when she also uses strategy $\kappa^\dagger$, that is,
$$
  u(\pi|\kappa^\dagger,\kappa^\dagger_{\neg n})
  =  \Exp^{(\kappa^\dagger,\kappa^\dagger_{\neg n})} \left[ \left. \int_0^\infty
         \left\{\astrut{2} [1-\kappa^\dagger(\pi_t)] s + \kappa^\dagger(\pi_t) m(\pi_t) - f(\pi_t) \right\}\,dt
         \ \right| \pi_0 = \pi \right].
$$
By definition, $u^*(\cdot|\kappa^\dagger_{\neg n}) \geq u(\cdot|\kappa^\dagger,\kappa^\dagger_{\neg n})$.
We shall establish the converse inequality via a comparison result for viscosity sub- and supersolutions.

We know that both functions are bounded.
Assume for now that they are actually continuous on $\Delta_L$; we will justify this assumption later.
While the following result and its proof are standard, we include them for the sake of a self-contained treatment.

\begin{lemma} \label{lem: subsolution}
The value function $u^*(\cdot|\kappa^\dagger_{\neg n})$ is a viscosity subsolution of the HJB equation.
\end{lemma}
\proof{
We simplify the notation by writing $u$ instead of $u^*(\cdot|\kappa^\dagger_{\neg n})$.

Consider $\phi \in C^2(\Delta_L)$ and $\pi_0 \in \overset{\circ}{\Delta}_L$ such that $u - \phi \leq 0 = u(\pi_0) - \phi(\pi_0)$.
To establish that $u$ is a viscosity subsolution of \eref{eq: HJB-1}, we must show that
$$
\max_{k \in [0,1]} \left\{ (1-k) s + k m(\pi_0) - f(\pi_0) + [k_0 + (N-1) \kappa^\dagger(\pi_0) + k] {\cal G}\phi(\pi_0) \right\} \geq 0.
$$
Suppose that this is not the case, so that
$$
(1-k) s + k m(\pi_0) - f(\pi_0) + [k_0 + (N-1) \kappa^\dagger(\pi_0) + k] {\cal G}\phi(\pi_0) < 0
$$
for all $k \in [0,1]$.
For $\varepsilon > 0$, define $\psi \in C^2(\Delta_L)$ by
$$
\psi(\pi) = \phi(\pi) + \varepsilon \| \pi - \pi_0 \|^4
$$
and note that $\psi \to \phi$ uniformly as $\varepsilon \to 0$.
For $\delta > 0$, let $B_\delta(\pi_0) \subset \R^L$ be the open ball of radius $\delta$ centered at $\pi_0$.
By continuity, we can find $\varepsilon, \delta > 0$ such that $B_\delta(\pi_0) \subset \overset{\circ}{\Delta}_L$ and
$$
(1-k) s + k m(\pi) - f(\pi) + [k_0 + (N-1) \kappa^\dagger(\pi) + k] {\cal G}\psi(\pi) < 0
$$
for all $k \in [0,1]$ and all $\pi \in B_\delta(\pi_0)$.
As $\pi_0$ is a strict maximizer of $u - \psi$, moreover, there exists $\gamma > 0$ such that $u(\pi) - \psi(\pi) \leq - \gamma$ for $\pi \in \Delta_L \setminus B_\delta(\pi_0)$.
Suppose now that player $n$ uses the strategy $\kappa \in \cal K$ against the other players' common strategy $\kappa^\dagger$.
Define $\tau = \inf \{t>0\!: \|\pi_t - \pi_0\| > \delta\}$.
As $k_0 >0$, we have $\Exp^{(\kappa,\kappa^\dagger_{\neg n})} [\tau] < \infty$ and
\begin{eqnarray*}
\lefteqn{\Exp^{(\kappa,\kappa^\dagger_{\neg n})} \left[ \int_0^\tau
         \left\{\astrut{2} [1-\kappa(\pi_t)] s + \kappa(\pi_t) m(\pi_t) - f(\pi_t) \right\}\,dt
         + u(\pi_\tau)
         \right] - u(\pi_0)} \\
& \leq & \Exp^{(\kappa,\kappa^\dagger_{\neg n})} \left[ \int_0^\tau
         \left\{\astrut{2} [1-\kappa(\pi_t)] s + \kappa(\pi_t) m(\pi_t) - f(\pi_t) \right\}\,dt
         + \psi(\pi_\tau)
         \right] - \psi(\pi_0) - \gamma \\
& = & \Exp^{(\kappa,\kappa^\dagger_{\neg n})} \left[ \int_0^\tau
         \left\{\astrut{2} [1-\kappa(\pi_t)] s + \kappa(\pi_t) m(\pi_t) - f(\pi_t) + [k_0 + (N-1) \kappa^\dagger(\pi) + \kappa(\pi_t)] {\cal G}\psi(\pi_t)\right\}\,dt
         \right] - \gamma \\
& < & - \gamma,
\end{eqnarray*}
where the equality in the third line follows from Dynkin's formula.
But this contradicts the dynamic programming principle, which states that
$$
u(\pi_0) = \sup_{\kappa \in \cal K} \Exp^{(\kappa,\kappa^\dagger_{\neg n})} \left[ \int_0^\tau
         \left\{\astrut{2} [1-\kappa(\pi_t)] s + \kappa(\pi_t) m(\pi_t) - f(\pi_t) \right\}\,dt
         + u(\pi_\tau)
         \right].
\vspace{-3ex}
$$
}

\begin{lemma} \label{lem: supersolution}
The payoff function $u(\cdot|\kappa^\dagger,\kappa^\dagger_{\neg n})$ is a viscosity supersolution of the HJB equation.
\end{lemma}
\proof{
We simplify the notation by writing $u$ instead of $u(\cdot|\kappa^\dagger,\kappa^\dagger_{\neg n})$.

Consider $\phi \in C^2(\Delta_L)$ and $\pi_0 \in \overset{\circ}{\Delta}_L$ such that $u - \phi \geq 0 = u(\pi_0) - \phi(\pi_0)$.
For any deterministic time $\tau > 0$,
\begin{eqnarray*}
0
& = & \Exp^{(\kappa^\dagger,\kappa^\dagger_{\neg n})} \left[ \int_0^\tau
         \left\{\astrut{2} [1-\kappa^\dagger(\pi_t)] s + \kappa^\dagger(\pi_t) m(\pi_t) - f(\pi_t) \right\}\,dt
         + u(\pi_\tau)
         \right] - u(\pi_0) \\
& \geq & \Exp^{(\kappa^\dagger,\kappa^\dagger_{\neg n})} \left[ \int_0^\tau
         \left\{\astrut{2} [1-\kappa^\dagger(\pi_t)] s + \kappa^\dagger(\pi_t) m(\pi_t) - f(\pi_t) \right\}\,dt
         + \phi(\pi_\tau)
         \right] - \phi(\pi_0) \\
& = & \Exp^{(\kappa^\dagger,\kappa^\dagger_{\neg n})} \left[ \int_0^\tau
         \left\{\astrut{2} [1-\kappa^\dagger(\pi_t)] s + \kappa^\dagger(\pi_t) m(\pi_t) - f(\pi_t) + [k_0 + N \kappa^\dagger(\pi_t)] {\cal G}\phi(\pi_t) \right\}\,dt
         \right]
\end{eqnarray*}
by Dynkin's formula.
Dividing through by $\tau$ and letting $\tau \to 0$, we get
$$
[1-\kappa^\dagger(\pi_0)] s + \kappa^\dagger(\pi_0) m(\pi_0) - f(\pi_0) + [k_0 + N \kappa^\dagger(\pi_0)] {\cal G}\phi(\pi_0) \leq 0,
$$
which is equivalent to
$$
\frac{[k_0 + (N-1) \kappa^\dagger(\pi_0)] [s - m(\pi_0)] - [f(\pi_0) - s]}{k_0 + N \kappa^\dagger(\pi_0)} - [s - m(\pi_0)] + {\cal G}\phi(\pi_0) \leq 0.
$$
As
$$
\kappa^\dagger(\pi_0) \in \arg\max_{k \in [0,1]}
    \frac{[k_0 + (N-1) \kappa^\dagger(\pi_0)] [s - m(\pi_0)] - [f(\pi_0) - s]}{k_0 + (N-1) \kappa^\dagger(\pi_0) + k},
$$
$u$ is thus a viscosity supersolution of \eref{eq: HJB-2}.
}

The comparison result that yields the inequality $u^*(\cdot|\kappa^\dagger_{\neg n}) \leq u(\cdot|\kappa^\dagger,\kappa^\dagger_{\neg n})$ is due to Ishii and Yamada (1993).
These authors consider functional equations $F(x,u,Du,D^2u,u-Mu) = 0$ such that
$$
F(x,r,p,X+Y,d) \leq F(x,r,p,X,d+q)
$$
for all $(x,r,p,X,d) \in \Omega \times \R \times \R^L \times \S^L \times \R$, all positive semidefinite $Y \in \S^N$ and all $q \geq 0$.
This means that $F$ corresponds to $-H$ here.\footnote{
Note that Ishii and Yamada (1993) allow the value of the solution to enter as a separate variable besides its difference with the nonlocal operator.
Because of the absence of discounting, this generality is not needed here, so $H$ has one argument fewer.}
As a consequence, the inequalities defining sub- and supersolutions in terms of $F$ are the opposite of those in terms of $H$.

There is a second, more substantive difference between the definitions of Ishii and Yamada (1993) and ours.
Translated back into our setting, a function $u \in C(\overline{\Omega})$ is a viscosity subsolution of \eref{eq: interior} in their sense
if for every $\phi \in C^2(\Omega)$ and every $x_0 \in \Omega$ such that $\phi-u$ has a local minimum in $x_0$,
$$
H(x_0,D\phi(x_0),D^2\phi(x_0),u(x_0)-Mu(x_0)) \geq 0.
$$
Analogously,
a function $u \in C(\overline{\Omega})$ is a viscosity supersolution of \eref{eq: interior} in their sense
if for every $\phi \in C^2(\Omega)$ and every $x_0 \in \Omega$ such that $\phi-u$ has a local maximum at $x_0$,
$$
H(x_0,D\phi(x_0),D^2\phi(x_0),u(x_0)-Mu(x_0)) \leq 0.
$$
In these alternative definitions, therefore, $u$ is replaced by $\phi$ only as far as the gradient and Hessian are concerned, but not in the nonlocal term.
When $M$ is an integral operator of the type considered here, however, an argument in Alvarez and Tourin (1996, p.~300) implies that these definitions are in fact equivalent to ours.\footnote{
See Azimzadeh, Bayraktar and Labahn (2018, Section 2) for a related discussion.}

\begin{lemma} \label{lem: comparison}
Let a function $v \in C(\partial \Delta_L)$ be given.
Suppose that $\underline u$ is a viscosity subsolution of the HJB equation,
$\overline u$ a viscosity supersolution,
and $\underline u \leq v \leq \overline u$ on $\partial \Delta_L$.
Then $\underline u \leq \overline u$ on $\Delta_L$.
\end{lemma}
\proof{
Equation \eref{eq: HJB-2} takes the form assumed in Ishii and Yamada (1993) with the domain $\Omega = \overset{\circ}{\Delta}_L$, the function
$$
F(x,p,X,d) =  - \frac{1}{2\sigma^2} R(x)' X R(x) + L(x)' p + \lambda(x) d  - c(x)
$$
where
$$
R(x) = \left( \begin{array}{c}
x_1 [\rho_1 - \rho(x)] \\
\vdots \\
x_L [\rho_L - \rho(x)]
\end{array} \right)
, \qquad L(x) = \left( \begin{array}{c}
x_1 [\lambda_1 - \lambda(x)] \\
\vdots \\
x_L [\lambda_L - \lambda(x)]
\end{array} \right)
$$
and
\begin{eqnarray*}
c(x)
& = & \max_{k \in [0,1]}
    \frac{[k_0 + (N-1) \kappa^\dagger(x)] [s - m(x)] - [f(x) - s]}{k_0 + (N-1) \kappa^\dagger(x) + k}
    - [s - m(x)] \\
& = & \astrut{4.5} \frac{[k_0 + (N-1) \kappa^\dagger(x)] [s - m(x)] - [f(x) - s]}{k_0 + N \kappa^\dagger(x)}
    - [s - m(x)],
\end{eqnarray*}
and the operator
$$
Mu(x) = \frac{1}{\lambda(x)} \int_{\Rnotz} u(j(x,h)) \,\nu(x)(dh).
$$
It is straightforward to check that $F$, $M$ and the function $B(x,u) = u - v(x)$ defined on $\partial \Delta_L \times \R$ satisfy all the conditions imposed by Ishii and Yamada (1993).
The result thus follows from their Theorem 3.1.
}

\begin{cor} \label{cor}
$u^*(\cdot|\kappa^\dagger_{\neg n}) = u(\cdot|\kappa^\dagger,\kappa^\dagger_{\neg n})$.
\end{cor}
\proof{
The proof is by induction over the dimension of the faces of the simplex.
The 0-faces (vertices) correspond to degenerate beliefs that assign probability 1 to one of the states;
at all these vertices, both functions assume the value 0.
An application of Lemma \ref{lem: comparison} for $L=1$ now yields
$u^*(\cdot|\kappa^\dagger_{\neg n}) = u(\cdot|\kappa^\dagger,\kappa^\dagger_{\neg n})$
along any 1-face (edge) of the simplex.
Applying the lemma for $L=2$ then proves this identity for all 2-faces (facets), and so on until the entire simplex is covered.
}

It remains to justify our assumption that the functions $u^*(\cdot|\kappa^\dagger_{\neg n})$ and $u(\cdot|\kappa^\dagger,\kappa^\dagger_{\neg n})$ are continuous.
In fact, using upper semicontinuous and lower semicontinuous envelopes, Ishii and Yamada (1993) define the notion of viscosity sub- and supersolution for functions that are merely locally bounded.
Lemmas \ref{lem: subsolution} and \ref{lem: supersolution} still hold then, and Lemma \ref{lem: comparison} generalizes in a way that ensures that any viscosity solution satisfying a continuous boundary condition must be continuous overall; see Ishii and Yamada (1993, Corollary 3.3).
Continuity of the functions in question follows from an iterative application of this result as in the proof of Corollary \ref{cor}.


\subsubsection*{Verification that $\kappa^\dagger \in \cal K$ for Brownian Payoffs and Normal Prior}

From the main body of the text, for $m < s$ we have
$$
  I(\pi)
= \Phi(z) - 1 + z^{-1} \phi(z)
$$
where $z = (s-m) \tau^{1/2}$.

The function $F(z) = \Phi(z) - 1 + z^{-1} \phi(z)$ is a strictly decreasing bijection from $]0,\infty[$ to itself with first derivative $F'(z) = - z^{-2} \phi(z)$.
For any positive real number $c$, therefore, we have $I(\pi) = c$ if and only if $(s-m) \tau^{1/2} = F^{-1}(c)$.
At any such $(m,\tau)$ in the half-plane $\Pi = \R \, \times \, [\tau_0,\infty[\,$, we have
$\partial I/\partial m = - F'(F^{-1}(c)) \, \tau^{1/2}$
and
$\partial I/\partial \tau = \half F'(F^{-1}(c)) F^{-1}(c)\, \tau^{-1}$.

To verify that $\kappa^\dagger \in \cal K$, it suffices to show that $I \tau^{-1}$ is Lipschitz continuous on
$\Pi(a,b) = \left\{ \pi \in \Pi: a \leq I(\pi) \leq b \right\}$
for any positive real numbers $a < b$.
For $I(\pi) = c$, we have
$\partial (I \tau^{-1})/\partial m = - F'(F^{-1}(c)) \, \tau^{-1/2}$
and
$\partial (I \tau^{-1})/\partial \tau = \left( \half F'(F^{-1}(c)) F^{-1}(c) - c \right) \tau^{-2}$.
This establishes that both partial derivatives of $I \tau^{-1}$ are bounded along any level curve $I(\pi)=c$ in $\Pi$.
Letting $c$ range from $a$ to $b$ shows that they are bounded on the whole of $\Pi(a,b)$, so $I \tau^{-1}$ is indeed Lipschitz continuous there.


\subsubsection*{Verification that $\kappa^\dagger \in \cal K$ for Poisson Payoffs and Gamma Prior}

Again from the main body of the text, for $m(\pi) = \alpha/\beta < s$ we have
$$
  I(\pi)
= \frac{s\,G(s; \alpha, \beta) - \divn{\alpha}{\beta}\,G(s; \alpha+1, \beta)} {s-\divn{\alpha}{\beta}} - 1.
$$
We fix $\alpha$ as well as positive real numbers $a < b$.
To verify that $\kappa^\dagger \in \cal K$, it suffices to show that $I(\alpha,\cdot)$ is Lipschitz continuous on the set $B(a,b)= \left\{ \beta \in \ ]\divn{\alpha}{s},\infty[ \ : a \leq I(\pi) \leq b \right\}$.
To this end, we note first that
$$
G(s; \alpha, \beta) - G(s; \alpha+1, \beta)
= \int_0^s \frac{\beta^\alpha}{\Gamma(\alpha)} \, x^{\alpha-1} e^{-\beta\mu} \left[ 1 - \frac{\beta\mu}{\alpha} \right] \, d\mu.
$$
For $\beta = {\alpha}/{s}$ and $\mu < s$, the term in square brackets under the integral is positive, so we have
$G(s; \alpha, \divn{\alpha}{s}) - G(s; \alpha+1, \divn{\alpha}{s}) > 0$.
For $\beta \searrow \divn{\alpha}{s}$, therefore, the numerator $s\,G(s; \alpha, \beta) - \divn{\alpha}{\beta}\,G(s; \alpha+1, \beta)$ in the above expression for $I(\pi)$ tends to a positive limit.
Given that $I(\pi)$ is finite for $\beta \in B(a,b)$,
this implies that the denominator in the above expression must be bounded away from 0,
i.e.\
$\beta$ must be bounded away from ${\alpha}/{s}$ on $B(a,b)$.
Using the fact that
$$\frac{\partial G(s; \alpha, \beta)}{\partial \beta} = \frac{\alpha}{\beta} \left[ G(s; \alpha, \beta) - G(s; \alpha+1, \beta) \right],$$
it is now straightforward to verify that $I(\alpha,\cdot)$ has a bounded first derivative on $B(a,b)$.

\AppendixOut               

\newpage


\def\next{\noindent \hangindent=3em \hangafter 1}

\parskip 0.5ex
\setstretch{1.05}

\def \ema  {Econometrica}
\def \jeea {Journal of the European Economic Association}
\def \jet  {Journal of Economic Theory}
\def \jf   {Journal of Finance}
\def \jpe  {Journal of Political Economy}
\def \rand {RAND Journal of Economics}
\def \res  {Review of Economic Studies}
\def \snk  {Sankhy\={a}}
\def \te   {Theoretical Economics}

\newcommand{\AJ}[6]{\next {#1}
(#2): ``#3," {\it #4}, {\bf #5}, #6.}

\newcommand{\ajo}[6]{\AJ{\sc #1}{#2}{#3}{#4}{#5}{#6}}

\newcommand{\ajt}[7]{\AJ{{\sc #1 and #2}}{#3}{#4}{#5}{#6}{#7}}

\newcommand{\AB}[6]{\next {#1}
(#2): ``#3," in #4, {\it #5}, #6.}

\newcommand{\abo}[6]{\AB{\sc #1}{#2}{#3}{#4}{#5}{#6}}

\newcommand{\abt}[7]{\AB{{\sc #1 and #2}}{#3}{#4}{#5}{#6}{#7}}

\newcommand{\BK}[4]{\next {#1}
(#2): {\it #3}. #4.}

\newcommand{\bko}[4]{\BK{\sc #1}{#2}{#3}{#4}}

\newcommand{\bkt}[5]{\BK{{\sc #1 and #2}}{#3}{#4}{#5}}

\newcommand{\WP}[4]{\next {#1}
(#2): ``#3," #4.}

\newcommand{\wpo}[4]{\WP{\sc #1}{#2}{#3}{#4}}

\newcommand{\wpt}[5]{\WP{{\sc #1 and #2}}{#3}{#4}{#5}}

\section*{References}

\ajt{Alvarez, O.}{A.\ Tourin} {1996}
{Viscosity Solutions of Nonlinear Integro-Differential Equations}
{Annales de l'Institut Henri Poincar\'{e}} {13} {293--317}

\ajt{Azimzadeh, P., E.\ Bayraktar}{G.\ Labahn} {2018}
{Convergence of Implicit Schemes for Hamilton-Jacobi-Bellman Quasi-Variational Inequalities}
{SIAM Journal on Control and Optimization} {56} {3994--4016}

\ajt{Bergemann, D.}{J.\ V\"{a}lim\"{a}ki} {1997}
{Market Diffusion with Two-sided Learning}
{\rand} {28} {773--795}

\ajt{Bergemann, D.}{J.\ V\"{a}lim\"{a}ki} {2002}
{Entry and Vertical Differentiation}
{\jet} {106} {91--125}

\abt{Bergemann, D.}{J.\ V\"alim\"aki} {2008}
{Bandit Problems}
{S.\ Durlauf and L.\ Blume (eds.)}
{The New Palgrave Dictionary of Economics (Second Edition)}
{Basingstoke and New York: Palgrave Macmillan}

\abt{Blum, A.}{Y.\ Mansour} {2007}
{Learning, Regret Minimization, and Equilibria}
{N.\ Nisan, T.\ Roughgarden, E.\ Tardos, V.\ Vazirani (eds.)}
{Algorithmic Game Theory}
{pp.\ 79-102, Cambridge: Cambridge University Press}

\ajt{Bolton, P.}{C.\ Harris} {1999}
{Strategic Experimentation}
{\ema} {67} {349--374}

\abt{Bolton, P.}{C.\ Harris} {2000}
{Strategic Experimentation: the Undiscounted Case}
{P.J.\ Hammond and G.D.\ Myles (eds.)}
{Incentives, Organizations and Public Economics -- Papers in Honour of Sir James Mirrlees}
{pp.\ 53--68, Oxford: Oxford University Press}

\ajo{Bonatti, A.} {2011}
{Menu Pricing and Learning}
{American Economic Journal: Microeconomics} {3} {124--163}

\ajt{Bubeck, S.}{N.\ Cesa-Bianchi} {2012}
{Regret Analysis of Stochastic and Nonstochastic Multi-Armed Bandit Problems}
{Foundations and Trends in Machine Learning} {5} {1--122}

\ajo{Chernoff, H.} {1968}
{Optimal Stochastic Control}
{\snk} {30} {221--252}

\ajt{Cohen, A.}{E.\ Solan} {2013}
{Bandit Problems with L\'{e}vy Payoff Processes}
{Mathematics of Operations Research} {38} {92--107}

\bko{DeGroot, M.} {1970}
{Optimal Statistical Decisions}
{New York: McGraw Hill}

\ajt{Ding, Z.}{I.O.\ Ryzhov} {2016}
{Optimal Learning with Non-Gaussian Rewards}
{Advances in Applied Probability} {48} {112--136}

\ajo{Dutta, P.K.} {1991}
{What Do Discounted Optima Converge to?: A Theory of Discount Rate Asymptotics in Economic Models}
{\jet} {55} {64--94}

\bko{Dynkin, E.B.} {1965}
{Markov Processes, Vol.\ I}
{Berlin: Springer}

\abt{Engelbert, H.J.}{W.\ Schmidt} {1984}
{On One-Dimensional Stochastic Differential Equations with Generalized Drift}
{M.\ M\'{e}tivier, E.\ Pardoux (eds.)}
{Lecture Notes in Control and Information Sciences}
{vol.\ 69, pp.\ 143--155, Berlin: Springer}

\wpo{Harris, C.} {1988}
{Dynamic Competition for Market Share: An Undiscounted Model}
{Discussion Paper No.~30, Nuffield College, Oxford}

\wpo{Harris, C.} {1993}
{Generalized Solutions to Stochastic Differential Games in One Dimension}
{Industry Studies Program Discussion Paper No.~44, Boston University}

\abt{H\"{o}rner, J.}{A.\ Skrzypacz} {2016}
{Learning, Experimentation and Information Design}
{B.\ Honor\'{e}, A.\ Pakes, M.\ Piazzesi, L.\ Samuelson (Eds.)}
{Advances in Economics and Econometrics: Eleventh World Congress \rm{(Econometric Society Monographs), pp. 63-98}}
{Cambridge: Cambridge University Press}

\ajt{Ishii, K.}{N.\ Yamada} {1993}
{Viscosity Solutions of Nonlinear Second Order Elliptic PDEs Involving Nonlocal Operators}
{Osaka Journal of Mathematics} {30} {439--455}

\ajo{Jovanovic, B.} {1979}
{Job Matching and the Theory of Turnover}
{\jpe} {87} {972--990}

\bkt{Karatzas, I.}{S.E.\ Shreve} {1988}
{Brownian Motion and Stochastic Calculus}
{New York: Springer-Verlag}

\ajt{Ke, T.T.}{J.M.\ Villas-Boas} {2019}
{Optimal Learning Before Choice}
{\jet} {180} {383--437}

\ajt{Keller, G.}{S.\ Rady} {1999}
{Optimal Experimentation in a Changing Environment}
{\res} {66} {475--507}

\ajt{Keller, G.}{S.\ Rady} {2003}
{Price Dispersion and Learning in a Dynamic Differentiated-Goods Duopoly}
{\rand} {34} {138--165}

\ajt{Keller, G.}{S.\ Rady} {2010}
{Strategic Experimentation with Poisson Bandits}
{\te} {5} {275--311}

\ajt{Keller, G.}{S.\ Rady} {2015}
{Breakdowns}
{\te} {10} {175--202}

\ajt{Keller, G., S.\ Rady}{M.\ Cripps} {2005}
{Strategic Experimentation with Exponential Bandits}
{\ema} {73} {39--68}

\ajo{Lai, T.L.} {1987}
{Adaptive Treatment Allocation and the Multi-Armed Bandit Problem}
{Annals of Statistics} {15} {1091--1114}

\bkt{Liptser, R.S.}{A.N.\ Shiryayev} {1977}
{Statistics of Random Processes I}
{New York: Springer-Verlag}

\ajt{Moscarini, G.}{F.\ Squintani} {2010}
{Competitive Experimentation with Private Information: The Survivor's Curse}
{\jet} {145} {639--660}

\bkt{{\O}ksendal, B.}{A.\ Sulem} {2007}
{Applied Stochastic Control of Jump Diffusions ($2^{\text{nd}}$ Edition)}
{New York: Springer-Verlag}

\ajt{Peitz, M., S.\ Rady}{P.\ Trepper} {2017}
{Experimentation in Two-Sided Markets}
{\jeea} {15} {128--172}

\bko{Pham, H.} {2009}
{Continuous-time Stochastic Control and Optimization with Financial Applications}
{New York: Springer-Verlag}

\ajo{Ramsey, F.P.} {1928}
{A Mathematical Theory of Savings}
{Economic Journal} {38} {543--559}

\ajo{Trotter, H.F.} {1959}
{On the Product of Semi-Groups of Operators}
{Proceedings of the American Mathematical Society} {10} {545--551}

\ajo{Veronesi, P.} {2000}
{How Does Information Quality Affect Stock Returns?}
{\jf} {55} {807--837}

\end{document}